\documentclass[english]{article}

\usepackage[T1]{fontenc}
\usepackage[utf8]{inputenc}
\usepackage[english]{babel}
\usepackage{authblk}
\makeatletter
\renewcommand\AB@affilsepx{\\}
\renewcommand\AB@affilnote[1]{{\textsuperscript{#1}}}
\makeatother

\usepackage{amsmath,mathtools,amssymb,amsthm,bm}
\usepackage{geometry}
\usepackage{xcolor}
\usepackage{comment}
\usepackage{natbib}
\usepackage{booktabs}
\usepackage{graphicx}
\usepackage{caption}
\usepackage{subfig}
\usepackage{float}
\usepackage{multirow}
\usepackage{tabularx,longtable,threeparttable}

\usepackage{algorithm,algcompatible,algpseudocode}
\newcommand*{\bsbeta}{\boldsymbol{\beta}}
\newcommand*{\bbR}{\mathbb{R}}
\newcommand*{\bbE}{\mathbb{E}}
\newcommand*{\mcalA}{\mathcal{A}}
\newcommand*{\mcalN}{\mathcal{N}}
\newcommand*{\mcalI}{\mathcal{I}}
\definecolor{Gray}{gray}{0.9}
\definecolor{White}{gray}{1}

\usepackage[unicode=true,bookmarks=false,breaklinks=false,
  pdfborder={0 0 1},colorlinks=false]{hyperref}
\hypersetup{colorlinks,
  linkcolor=red,anchorcolor=blue,citecolor=blue,urlcolor=blue}

\algnewcommand\algorithmicpara{\textbf{estimate in parallel for}}
\algblockdefx[PARA]{Para}{EndPara}[1]{\algorithmicpara\ #1}{\algorithmicend}
\algnewcommand\algorithmicdivide{\textbf{divide}}
\algnewcommand\algorithmicextract{\textbf{matrix decomposition}}
\algblockdefx[DIVIDE]{Divide}{EndDivide}{\algorithmicdivide}{\algorithmicend}
\algblockdefx[EXTRACT]{Extraction}{EndExtract}{\algorithmicextract}{\algorithmicend}

\algnewcommand{\algorithmicendif}{\textbf{end}}
\algblockdefx[IF]{If}{EndIf}[1]{\algorithmicif\ #1\ \algorithmicthen}{\algorithmicendif}

\newtheorem{theorem}{Theorem}[section]
\newtheorem{lemma}{Lemma}
\newtheorem{condition}{Condition}[section]
\newtheorem{remark}{Remark}[section]
\newtheorem{proposition}[theorem]{Proposition}
\newtheorem{corollary}[theorem]{Corollary}

\def\A{{\bf A}}

\def\0{{\bf 0}}

\def\calA{{\mathcal{A}}}
\def\calI{{\mathcal{I}}}
\def\bbeta{{\boldsymbol{\beta}}}
\def\hbeta{{\hat{\boldsymbol{\beta}}}}
\allowdisplaybreaks
\newcommand{\TableSpace}{%
  \par
  \ifdim\dimexpr\pagegoal-\pagetotal\relax<8\baselineskip
    \newpage
  \fi
}

\newcommand{\HP}{HP}
\newcommand{\GICstar}{\texorpdfstring{GIC\textsuperscript{*}}{GIC*}}
\newcommand{\HPBIC}{HP-BIC}
\newcommand{\HPGIC}{HP-\GICstar}
\newcommand{\SBM}{S$\beta$M}
\newcommand{\SBMBIC}{\SBM-BIC}
\newcommand{\SBMGIC}{\SBM-\GICstar}
\newcommand{\proposedcell}[1]{{\bfseries\boldmath #1}}

\title{Homogeneous pursuit in $\beta$ model}

\author{%
  Ergan Shang\textsuperscript{1}\thanks{To whom correspondence should be addressed. E-mail: eshang@andrew.cmu.edu.} \quad
  Si Chen\textsuperscript{2} \quad
  Canhong Wen\textsuperscript{2}%
}
\date{}

\begin{document}

\maketitle

\footnotetext[1]{Department of Statistics and Data Science, Carnegie Mellon University, Pittsburgh, PA 15213.}
\footnotetext[2]{Department of Statistics and Finance, School of Management, University of Science and Technology of China, Hefei, Anhui Province, China.}
\footnotetext[3]{The authors declare no conflict of interest.} 
\begin{abstract}

Data in forms of networks has been encountered in various prevalent fields at the arrival of big data era. To unveil and recover the true model with core units with numerous links, the $\beta$-model under sparsity constraint is proposed to meet the expectation of statisticians and engineers. Not only does this network model inherit spectacular property from the Erd\H{o}s-R\'enyi model, but also exhibits degree heterogeneity through tuning sparsity parameter $s$. In this paper, we propose Homogeneous Pursuit (\HP), an $\ell_0$-constrained maximum likelihood estimation method implemented through a primal-dual algorithm. To ease the harshness of computation, we find the surrogate function of the corresponding objective function to derive Majorization Minimization algorithm through the primal-dual process. Moreover, after meticulous evaluation, the algorithm brought about captures the property of fine convergence rate in a contraction mapping way. Under some mild conditions, the proposed generalized information criterion (\GICstar) for selecting the sparsity level is consistent to find the truth. Combining \HP{} with \GICstar{} gives the adaptive procedure \HPGIC; \HPBIC{} denotes the same estimation method with sparsity selected by BIC. Finally, simulation studies show that \GICstar{} improves support recovery relative to BIC in the reported comparisons, while \HP{} achieves lower computational cost while maintaining estimation accuracy comparable to the sparse benchmark methods. A microfinance application illustrates the use of \HPGIC{} for studying the association between network prominence and participation, with improved fit in the reported village-level regressions.

\end{abstract}

\setcounter{tocdepth}{2}


\section{Introduction}

The rapid growth of high-dimensional data has made understanding interactions among individual units increasingly important for scientists, engineers, and statisticians. Networks provide a natural framework for representing these interactions, with nodes corresponding to individual units and edges describing their relationships. This perspective has been widely adopted in science \citep{barabasi2013network,zhou2026asi}, biology \citep{zhang2025genetic,shang2025predicting}, social sciences \citep{shang2026inference}, economics \citep{pakes2017advances}, and communication \citep{adamic2005political}. Beyond these traditional applications, network-based representations also play an increasingly important role in deep learning, large language model benchmarking \citep{phan2025humanity,li2026long,shang2026llm,wu2026benchmark}, and recommendation systems \citep{shang2026erase}, capturing relationships among data instances, comparisons among models \citep{liu2026oscar}, and interactions between users and items. These applications highlight the importance of developing effective statistical methods for analyzing network structure and node heterogeneity.

In the early stage of the networks, probably the earliest and also the simplest network model attained attention is the Erd\H{o}s-R\'enyi model \citep{erdos59a,erdos1960evolution,gilbert1959random} which was arisen from combinatorics. In this earliest network model, all links can be described by only one parameter, where connections happened to different pairs of nodes via the same probability $p=\frac{e^\mu}{1+e^\mu}$. Although the birth of Erd\H{o}s-R\'enyi model came from mathematicians dedicating to pure Maths, its significance of statistical properties \citep{chen2021analysis} is still studied deeply by statisticians nowadays. In Chen's discussion \citep{chen2021analysis}, the estimation through maximum likelihood estimator(MLE)  displays meaningful characters such as consistency and asymptotically normaility for both dense and sparse networks.

Nonetheless, in practical scenarios, many real-world networks are armed with several `core' units or nodes, often more important to organize the whole structure than other `silent' nodes. With corn nodes included, more adges related to them are more prone to be established, exhibiting a higher probability for connection and high degrees. In spite of the spectacular characters of Erd\H{o}s-R\'enyi model, however, this model always reflects a state of \textit{homophily} \citep{mcpherson2001birds} phenomenon, that is, all nodes and edges share the same status instead of domination between each other, disconforming to the real world. With so many network models at hand, those equipped with distinguished features -- \textit{degree heterogeneity} \citep{barabasi2003scale} -- are always attached great significance to, attracting a great many statisticians devoting to illuminating the interactions. 

In order to impose degree heterogeneity property more specifically, many statistical models have been bred up. One of the most prominent examples is the $\beta$-model, named by Chatterjee \citep{chatterjee2011random}, using node-specific parameters explicitly \citep{britton2006generating}. Represented by an adjacency matrix $A=(A_{ij})_{1\leq i,j\leq n}$, an undirected and binary network of $n$ nodes is generated, with the connection probability\[ P(A_{ij}=1)=\frac{exp(\beta_i+\beta_j)}{1+exp(\beta_i+\beta_j)}. \]After Chatterjee gave the name of the model, consistency of MLE had been proved and also the estimation precision under $\ell_\infty$ norm \citep{chatterjee2011random} of order $\sqrt{\frac{logn}{n}}$. As a matter of fact, most networks nowadays containing certain level of degree heterogeneity usually have few high-degree core nodes with enormous links and a great many of points with few connections. 

Under the goal of unveilling the true model -- detecting both `core' nodes and `silent' points -- an urge of selecting which parameter to be exactly a fixed constant \citep{zhang20212},i.e. $\Vert\bsbeta-\bsbeta_0\mathbf{1}\Vert_0\leq s$ was arose which can be viewed as a generalization of Erd\H{o}s-R\'enyi model: when $\beta_j=\beta_i=\beta_0$, then $2\beta_0=\mu$. This constrain shrinks the gap between the reality and the theoretical models proposed, also illuminating the way aided by sparse optimization.

In the meantime, the $\beta$-model also branched into several new directions. For instance, undirected graphs have been extended into directed graphs with covariates and bipartite networks \citep{yan2019statistical,yan2016asymptotics}; another network model, the Sparse $\beta$-Model (\SBM) \citep{chen2021analysis} has also been brought up via a baseline $\mu$:\[ P(A_{ij}=1)=\frac{exp(\mu+\beta_i+\beta_j)}{1+exp(\mu+\beta_i+\beta_j)}, \]where $\mu\in\mathbb{R}$ and $\bsbeta\in\mathbb{R}^n_+$. With the constrain of $\Vert\bsbeta\Vert_0=s$, this model can also make the achievement of generalizing Erd\H{o}s-R\'enyi model; moreover the combination of both models mentioned can also portray an insightful statistical model \citep{stein2020sparse}.

Despite the splendid development history of the $\beta$-model in the past several decades, there remain some tough problems needed to be dealt with, both computationally and theoretically:
\begin{enumerate}
\item The first point is the estimation via statistical theories. Though the existence and consistency of MLE had been given at first \citep{chatterjee2011random}, nothing related to sparsity was considered then. Whereas a convenient and well-known monotonicity lemma \citep{chen2021analysis} was given, and yet every estimation was made on the condition that the active set is known in advance instead of computing and making recoveries of the set by efficient algorithms. As for other contributions related so far, complicated expressions and sets of troublesome conditions \citep{zhang20212} were exposited to be statisfied. Some even obtain the estimator through reparametrization, with the parameters left without figuring out the methods to tune \citep{kolaczyk2015question,krivitsky2011adjusting,chen2021analysis}. Either the sparsity constraint or the estimation problem was not paid attention to in the rest of the paper. Concentration of statistics often shifts to hypothesis testing \citep{mukherjee2018detection}, excess risk bounds \citep{chen2021analysis}, etc. 
\item The second major issue is computation. With the elegant statistical property of MLE at hand, the eagerness of deriving computationally tractable algorithms for estimating active set and parameters is always in urge. Although different types of penalized function have been implemented, such as $\ell_0$ norm \citep{chen2021analysis} and $\ell_2$ norm \citep{zhang20212}, a brief and concise convergence rate bound is rarely proved and presented. What's worse, many papers went forward to the estimation of $\bsbeta$, leading to a bad estimation for $\beta_0$ \citep{zhang20212}, let alone the parameters of interest. On the other hand, reasonable methodology for figuring out the tuning parameter is the vital ingrediant for good-performingly logical algorithms. Traditionally, information criterion such as AIC \citep{akaike1998information} and BIC \citep{schwarz1978estimating} are widely applied, but they work well mainly when the degree of dimension is fixed. Cross-validation may act as an aid, however, with the growth of dimensionality, AIC and cross-validation may fail to play such a role asymptotilcally \citep{zhang2010nearly,she2019cross}.
\end{enumerate}

We call our estimation method \emph{Homogeneous Pursuit} (\HP). Throughout the paper, \HPGIC{} denotes \HP{} with sparsity selected by our proposed criterion \GICstar, whereas \HPBIC{} uses BIC. For comparison, \SBMBIC{} and \SBMGIC{} apply BIC and \GICstar, respectively, to the Sparse $\beta$-Model estimation procedure of \citet{chen2021analysis}. Thus, the comparison separates the estimation procedure from the criterion used to select the sparsity level.

To address the weakness just mentioned in the $\beta$-model up to now, we made the following contributions:

\begin{enumerate}
\item First and foremost, we made the estimation of the parameters in the $\beta$-model, both the active set and $\bsbeta,\beta_0$. To cure the unbearable computational burden, we find the surrogate function of the target, converting the task into a convex function plus an $\ell_0$ penalization. Heading forward to the active set, a primal-dual algorithm was developed for \HP. Finally, thanks to the fine convergence rate of MM algorithm \citep{balakrishnan2017statistical}, the rationality of the algorithm as well as the estimator was founded.
\item Furthermore, to get over the pain points of classical information criterion, an avant-garde criterion called GIC \citep{nishii1984asymptotic} was proposed to incorporate them. The flexibility of this new criterion casts a bunch of light onto the road of finding reasonable tuning parameters. Within this existing GIC framework, we derive the specific criterion \GICstar{} and establish its consistency for selecting the sparsity level.
\end{enumerate}

This paper makes three main contributions. First, in Section~\ref{sec:model-algorithm}, we formulate \HP{} in the $\beta$-model as an $\ell_0$-constrained likelihood optimization problem and develop a primal-dual algorithm \citep{wen2020bess} using a surrogate function within the majorization-minimization (MM) framework \citep{lange2016mm}. In Section~\ref{sec:theory}, we establish convergence guarantees for the \HP{} algorithm and derive estimation error bounds under suitable conditions. Second, in Section~\ref{sec:selection}, we propose a generalized information criterion, \GICstar, for selecting the sparsity level and establish its model-selection consistency. Third, we evaluate \HPGIC{} through simulation studies and real-data analysis. The simulation results show improved support recovery with \HPGIC{} and lower reported computational cost with \HP{}, while maintaining estimation accuracy comparable to the sparse benchmarks. In the microfinance application, the estimated node effects are positively associated with household participation, suggesting that network prominence is relevant to microfinance adoption. These findings illustrate the practical value of the proposed framework for identifying heterogeneous nodes and relating their network positions to observed behavior.

\section{Model and Methodology}\label{sec:model-algorithm}

In this section, we first present our model setting and introduce the Homogeneous Pursuit (\HP) problem in the $\beta$-model and its estimation algorithms.

\subsection{Model setting}\label{data-model}

To begin with, we introduce some notations and definition in network studies. For a network with $n$ nodes, denote the set of nodes by $V$ and the set of edges by $E$, respectively. Then we can  represent a network by a graph $G_n=G_n(V,E)$ \citep{chen2021analysis}. Let $\A=(A_{ij})_{i,j=1}^n$ be the adjacency matrix with $A_{ij}\in\{0,1\}$ being an indicator whether nodes $i$ and $j$ are connected. That is, 
\[ 
A_{ij}=\begin{cases}
1 & \mathrm{if} \ \mathrm{nodes} \ \textit{$i$} \ \mathrm{and} \ \textit{$j$} \ \mathrm{are} \ \mathrm{connected}, \\
0 & \mathrm{if} \ \mathrm{nodes} \ \textit{$i$} \ \mathrm{and} \ \textit{$j$} \ \mathrm{are} \ \mathrm{not} \ \mathrm{connected}.
\end{cases}
\]
Throughout this paper, we consider undirected graphs with no self loops such that the adjacency matrix $\A$ is symmetric with zero diagonal entries. For node $i$, we define its degree by  $d_i=\sum_{j=1}^n A_{ij}$.

We consider the $\beta$-model proposed by \cite{chatterjee2011random}, which assumes that $A_{ij}$'s are independent Bernoulli random variables with
\[ 
P(A_{ij}=1)=\frac{e^{\beta_i+\beta_j}}{1+e^{\beta_i+\beta_j}}, 
\] 
where $\boldsymbol{\beta}=(\beta_1,\cdots,\beta_n)^\top\in\mathbb{R}^n$ is an unknown parameter. Note that $\beta_i$ measures the propensity of node $i$ to connect with other nodes, and the larger $\beta_i$ is, the more likely node $i$ is connected with other nodes. To estimate $\boldsymbol{\beta}$, we minimize the negative log-likelihood of this model, which is given by
\begin{equation}\label{eqn:LogLike}
\ell_n(\bsbeta)=-\sum_{i=1}^nd_i\beta_i+\sum_{1\leq i<j\leq n}\log(1+e^{\beta_i+\beta_j}).
\end{equation}

Despite these attractive properties, the $\beta$-model has a limitation when sparse networks are considered \citep{chen2021analysis}. In a sparse network, some of knots share the same connect with other nodes, that is, $\beta_i=\beta_0$ for some $i\in[n]$ and $\beta_0 \in \mathbb{R}$. To estimate $\boldsymbol{\beta}$ and $\beta_0$ with such constraint, we consider the following problem
\begin{equation}\label{eqn:model}
    \min_{\boldsymbol{\beta}\in\mathbb{R}^n}\ell_n(\bsbeta) \quad \mathrm{subject} \ \mathrm{to}\ \Vert\boldsymbol{\beta}-\beta_0\mathbf{1}\Vert_0\leq s ,
\end{equation}
where $\mathbf{1}$ is an $n$-dimensional vector with its elements being 1 and $s\in\{1,2,\cdots,n-1\}$ is an integer-value parameter to control the sparsity of the network. Denote the solution of \eqref{eqn:model} as $\hbeta$. When $s=n-1$, it is actually an unconstrained problem and the solution is the MLE of the $\beta$-model. When $s=1$, it reduces to 
\begin{equation}\label{eqn:er}
    P(A_{ij}=1)=\frac{e^{2\beta_0}}{1+e^{2\beta_0}}, 
\end{equation}
which coincides the Erd\H{o}s-R\'enyi model \citep{erdos59a,erdos1960evolution} with parameter $2\beta_0$. Here we are mainly interested in the case where $s<<n$, i.e., only $s$ knots have heterogeneous connect with others and the other $n-s$ knots share the same propensity to connect with other nodes.

The \SBM{} of \citet{chen2021analysis} uses nonnegative sparse node effects above a common baseline. In contrast, the \HP{} constraint in \eqref{eqn:model} allows a small set of nodes to deviate in either direction from $\beta_0$. We use \HPGIC{} for the adaptive procedure combining this estimator with \GICstar{} selection.

\subsection{Algorithm with fixed sparsity level}

Denote the support of $\bbeta-\beta_0\mathbf{1}$ as $\mathcal{A}$, that is, $\calA = \{i: \beta_i\neq \beta_0\}$. Assume that the cardinality of $\calA$ satisfies the constraint $|\calA|=s$ and let $\calI = \calA^c$. When $\calA$ is known, the problem \eqref{eqn:model} reduces to the following unconstrained problem
\begin{equation}
    \begin{aligned}
    \min & \bigg\{-\sum_{i\in\mcalA}d_i\beta_i-\sum_{i\in\mcalI}d_i\beta_0+\frac{1}{2}\sum_{i \neq j;i,j\in\mcalA} \log(1+e^{\beta_i+\beta_j})+ \\
    & +\vert\mcalI\vert\sum_{i\in\mcalA} \log(1+e^{\beta_i+\beta_0})+\frac{1}{2}\vert\mcalI\vert(\vert\mcalI\vert-1)\log(1+e^{2\beta_0})\bigg\} .
    \end{aligned}
\end{equation}
It is easily seen that the estimate for $\beta_0$ can be expressed as
\begin{equation}
    \beta_0=\frac{1}{2}\mathrm{logit}\left(\frac{\sum_{i\in\mcalI}d_i-\sum_{i\in\mcalA}d_i+\sum_{i,j\in\mcalA;i\neq j}\frac{e^{\beta_i+\beta_j}}{1+e^{\beta_i+\beta_j}}}{\vert\mcalI\vert(\vert\mcalI\vert-1)}\right) 
\end{equation}
where $\mathrm{logit}(p)$ is the logit function defined as $\mathrm{logit}(p) = \ln \frac{p}{1-p}$ for $p \in (0,1)$. When $\calA=\varnothing$, the estimate is simplified to
$$
\beta_0 = \frac1{2} \mathrm{logit}\left(\frac{\sum_{i=1}^n d_i}{n(n-1)}\right),
$$
which is also the MLE of $\beta_0$ in the Erd\H{o}s-R\'enyi model \eqref{eqn:er}. 

In practice, the support $\calA$ is usually unknown. In the following, we illustrate how to determine $\calA$. There are two barriers to solve the $\ell_0$-norm constrained maximum likelihood problem in \eqref{eqn:LogLike}. Firstly, even without the $\ell_0$-norm constraint, the minimization of $\ell_n(\bbeta)$ has no explicit solution and would involves expensive computational cost to iterative derive the solution. Secondly, the $\ell_0$-norm constrained optimization problem~\eqref{eqn:LogLike} is a combinatorial problem that seems difficult to solve.

To address these problems, at the $m$-th step, we introduce a surrogate function of $\ell_n(\bbeta)$ around the current estimate $\bbeta^m$ via a second order Taylor expansion. In specific, let $G^m=(G_1^m,\cdots,G_n^m)^T$ be the gradient of $\ell_n(\bsbeta)$ at $\bbeta^m$ with $G_i^m=-d_i+\sum_{j\neq i}\frac{\exp(\beta_i^m+\beta_j^m)}{1+\exp(\beta_i^m+\beta_j^m)}$. Then the surrogate function $S(\bsbeta|\bbeta^m)$ is defined by 
$$
S(\bsbeta|\bbeta^m) = \ell_n(\bbeta^m)+(G^m)^\top (\bsbeta-\bsbeta^m)+\frac{n-1}{4}\Vert\bsbeta-\bsbeta^m\Vert^2_2.
$$
Next lemma states that $S(\bsbeta|\bsbeta^m)$ is indeed a surrogate function of $\ell_n(\bsbeta)$ at $\bbeta^m$. The proof is deferred to \ref{App-B}.

\begin{lemma}\label{Lemma 1.}
The function $S(\bsbeta|\bsbeta^m)$ majorizes the objective function $\ell_n(\bsbeta)$ at $\bsbeta^m$. That is, 
%
%
\begin{align*} 
S(\bsbeta|\bsbeta^m) & \geq\ell_n(\bsbeta) \text{ for all }\ \bsbeta, \\
 S(\bsbeta^m|\bsbeta^m) & =\ell_n(\bsbeta^m).
\end{align*}

\end{lemma}

At the $m$-th step, we consider the following simplified optimization problem.
\begin{equation}\label{eqn:smodel}
    \min_{\bsbeta}S(\bsbeta|\bsbeta^m) \quad \mathrm{subject} \ \mathrm{to}\ \Vert\boldsymbol{\beta}-\beta_0\mathbf{1}\Vert_0\leq s.
\end{equation}
Denote the solution of the problem above by $\bsbeta^{m+1}$. Lemma~\ref{Lemma 1.} provides theoretical justification that a local optimum of the problem \eqref{eqn:model} would be guaranteed. This is because for all $m$, we have 
\[ 
\ell_n(\bsbeta^{m+1})\leq S(\bsbeta^{m+1}|\bsbeta^m)\leq S(\bsbeta^m|\bsbeta^m)=\ell_n(\bsbeta^m). 
\]
and the function $\ell_n$ has a finite lower bound 0.

Note that without the $\ell_0$-norm constraint, the optimization problem~\eqref{eqn:smodel} can be solved in a coordinate-wise way.  In particular, the solution is $\tilde{\bbeta}_j = \bbeta^m_j-\frac{2}{n-1} G^m_j$, $j=1,\dots, n$. To determine which elements of $\bsbeta^{m+1}$ should be equal to $\beta_0$, we consider the sacrifice of switching from $\tilde{\bbeta}_j$ to $\beta_0$ by fixing other coordinates, which is defined by
\begin{align*} 
    \Delta_j = & S\left((\bsbeta^{m+1}_1,\dots,\bsbeta^{m+1}_{j-1},\beta_0,\bsbeta^{m+1}_{j+1},\dots, \bsbeta^{m+1}_n)|\bsbeta^m\right) - \\
    & S\left((\bsbeta^{m+1}_1,\dots,\bsbeta^{m+1}_{j-1},\tilde{\bbeta}_j,\bsbeta^{m+1}_{j+1},\dots, \bsbeta^{m+1}_n)|\bsbeta^m\right)\\
    =& \frac{n-1}{4}\left( \beta_0 -{\bbeta}_j^{m}  +\frac{2}{n-1}G_j^m\right)^2.
\end{align*}
Let $\Delta_{[1]}\geq\cdots\geq\Delta_{[n]}$ denote the decreasing rearrangement of $\{\Delta_j: j=1,\dots,n\}$. Then for a pre-specified sparsity level $s$, we can determine the active set by
\[ 
\mcalA=\{j:\Delta_j\geq\Delta_{[s]}\}, 
\] 
and the inactive set by
\[ 
\mcalI=\{j:\Delta_j<\Delta_{[s]}\}.
\]

Now we summarize the arguments above in \textbf{Algorithm 1} below.

\begin{algorithm}[!ht]
	\caption{\HP: Majorization active set algorithm}
	\label{MM Algorithm}
	\begin{algorithmic}[1]
		\Require Adjacency matrix $A$,  sparsity level $s$, the maximum number of iterations $m_{\max}$, and tolerance $\epsilon$.
		\State Initialize $\beta_0^0 = \frac1{2} \mathrm{logit}\left(\frac{\sum_{i=1}^n d_i}{n(n-1)}\right)$, $\bbeta^0_i = \beta_0^0$ and $G_i^0=-d_i+\sum_{j\neq i}\frac{\exp(\beta_i^0+\beta_j^0)}{1+\exp(\beta_i^0+\beta_j^0)}$ for $i=1,\dots, n$.
            \State \textbf{For} $m\in\{0,1,2,\cdots,m_{\max}\}$ do
            \State \quad Determine $\Delta_j^{m+1}=\frac{n-1}{4}\left( \beta_0 -{\bbeta}_j^{m}  +\frac{2}{n-1}G_j^m\right)^2,\ j=1,\cdots,n$.
		   \State \quad Update the active and inactive sets by\[ \mcalA^{m+1}=\{j:\Delta_j^{m+1}\geq\Delta_{[s]}^{m+1}\},\quad \mcalI^{m+1}=(\mcalA^{m+1})^c .\]
            \State \quad Determine $\bsbeta^m$ by
            \[ 
            \bsbeta^{m+1}_{\mcalI^{m+1}}=\beta_0^{m} \quad \mathrm{and} \quad \beta^{m+1}_{\mcalA^{m+1}}=\left(\bsbeta^{m}-\frac{2}{n-1}G^{m} \right)_{\mcalA^{m+1}}.
            \]
            \State \quad Let \[
            p=\frac{\sum_{i\in\mcalI^{m+1}}d_i-\sum_{i\in\mcalA^{m+1}}d_i+\sum_{i\neq j, i,j \in\mcalA^{m+1} }\frac{\exp(\beta^{m+1}_i+\beta^{m+1}_j)}{1+\exp(\beta^{m+1}_i+\beta^{m+1}_j)}}{ (n-s)(n-s-1)}.
            \]
            \quad If $p< 1$, then $\beta_0^{m+1}=\frac1{2}\mathrm{logit}(p)$, else break.
		\State \quad \textbf{If} $\Vert\bsbeta^{m+1}-\bsbeta^m\Vert_2<\epsilon$, break.
		\State \textbf{end For}
		\State \Return $\hat{\bsbeta}=\bsbeta^{m+1}, \hat \mcalA =\mcalA^{m+1}$.
	\end{algorithmic}
\end{algorithm}

We stop the iteration according to whether $\Vert\bsbeta^{m+1}-\bsbeta^m\Vert_0$ is small enough(equal) or a maximum allowed iteration number is achieved.

\begin{remark}
 Note that the initialization is actually with empty active set. That is, the ER model.
\end{remark}

\subsection{Algorithm with tuned sparsity level}

In practise, the sparsity level is unknown and needs to be chosen. Here we search over a list of candidate values and determine an optimal sparsity level by minimizing some information criterion. To be specific, we use the proposed generalized information criterion (\GICstar), which is defined by

\begin{equation}\label{eqn:gic}
    \mathrm{GIC}^{*}(s) = 2\ell_n\left(\hat\bsbeta (s)\right) + s \log \left(\frac{n(n-1)}{2}\right)\cdot\log(\log n),
\end{equation}
where $\hat\bsbeta(s)$ is the output of \textbf{Algorithm 1} with sparsity level $s$.

Following \cite{stein2020sparse}, we search over an unequal sequential list of sparsity levels determined by the distinctive values of degrees in the network. In particular, denote by $m$ the number of distinctive values in the degree sequence $\{d_i, i=1,\dots, n\}$. Then let $d_{(1)} > d_{(2)} > \cdots > d_{(m)}$ be the ordered distinctive values of $d_i$’s. For each $k \in \{1,2,\cdots,m\}$, denote the set of indices of those $d_i$'s equal to $d_{(k)}$ by $S_k =\left\{i \in \left\{1,\dots, n \right\}:d_i = d_{(k)}\right\}$ and its cardinality by $s_k=|S_k|$. We can determine an optimal sparsity level by searching over the following list $\{ s_1,s_1+s_2, \dots,\sum_{k=1}^{m}s_k \}$.

When the number of distinctive degrees is large or a sparse network is believed, we can stop the search early by setting the largest sparsity level by $s_{\max}$. Let $m_s = \arg\max_{1\leq k \leq m}\{\sum_{j=1}^k s_j \leq s_{\max} \}$, then the sequential list we search is $\{ s_1,s_1+s_2, \dots,\sum_{k=1}^{m_s}s_k \}$.

\begin{algorithm}[!htb]
	\caption{\HPGIC: Adaptive majorization active set algorithm}
	\label{Algorithm 2}
	\begin{algorithmic}[1]
		\Require Adjacency matrix $A$, the maximum support size $s_{\max}$, and the other parameters for \textbf{Algorithm 1} ( the maximum number of iterations $m_{\max}$, and tolerance $\epsilon$).
		\State Set degree sequence $\mathbf{d}=(d_1,\cdots,d_n)^{T}$.
        \State Let $d_{(1)} > d_{(2)} > \cdots > d_{(m)}$ be the distinctive values of $d_i$’s, where $m$ is the number of distinctive values in $\{d_i, i=1,\dots, n\}$.
        \State Determine $S_k =\left\{i \in \left\{1,\dots, n\right\}:d_i = d_{(k)}\right\}$ and $s_k=|S_k|$, $k=1,2,\cdots,m$. Let $m_s = \arg\max_{1\leq k \leq m}\{\sum_{j=1}^k s_j \leq s_{\max} \}$.
		\State \textbf{For} $s=s_1,s_1+s_2,\cdots,\sum_{k=1}^{m_s}s_k$ \textbf{do}
		   \State \quad Run \textbf{Algorithm 1} with $A$, $s$, $m_{\max}$ and $\epsilon$. Denote the output as $(\hat \bsbeta_s,\hat \mcalA_s)$.
                \State \quad Calculate the \GICstar{} value $\mathrm{GIC}^{*}(s)$ with $\hat \bsbeta_s$ according to \eqref{eqn:gic}
		\State\textbf{end For}
		\State Determine the optimal sparsity level by minimizing the \GICstar{} values, i.e., 
            \[ 
                s_{opt}=\arg\min_s \mathrm{GIC}^{*}(s) .
            \]
		\State \Return $(s_{opt}, \hat \bsbeta_{s_{opt}},\hat \mcalA_{s_{opt}})$.
	\end{algorithmic}
\end{algorithm} 
\section{Theoretical properties}\label{sec:theory}

In this section, we investigate the theoretical properties of the MM algorithm used by \HP. We first give a mild condition on the estimation of $\beta_0$ through the algorithms and then show the distance between the iteration result and the MLE.

\begin{condition}\label{cond.1}
   Considering the ordered degree sequence $d_{[1]}\geq\cdots\geq d_{[n]}$ and $\mcalA^{max}=\{i:d_i\geq d_{[s]}\},$ we discuss our convergence rate estimation problem on the network such that \[O\left(n^{2-\rho}\right)\geq\sum_{i=1}^nd_i \ \mathrm{and} \ \sum_{i=1}^nd_i-2\sum_{i\in\mcalA^{max}}d_i\geq cn^{\frac{3}{2}+c'}\] for some positive $\rho$, $c$ and $c'$ which will cause no contradiction.
\end{condition}

It is natural to understand the first condition because the maximal sum of degree for a network is $O(n^2)$. In a network with several pivotal nodes and lots of "silent" nodes, it is reasonable to constrain sum of degree to have a upper bound $n^{2-\rho}$. Recall the definition of $p$ mentioned in the \textbf{Algorithm} \ref{MM Algorithm}, we can express $\sum_{i\in\mcalI^{m+1}}d_i-\sum_{i\in\mcalA^{m+1}}d_i=\sum_{i=1}^nd_i-2\sum_{i\in\mcalA^{m+1}}d_i$ in the numerator. The second condition describes that the difference between the sum of degrees of $n-s$ nodes with smaller degrees and the sum of degrees of $s$ nodes with the largest degrees reaches the order of more than $n^{\frac{3}{2}}$, expliciting a low sparsity level of $s$. Moreover, in the view of difficulty degree of conditional verification, we can calculate the degree of a network upon having obtained the data.

\subsection{Main results}
\begin{theorem}\label{Theorem 2.}
Denote by $\bsbeta^{m+1}$ the output at the $\mathrm{(m+1)}$-th iteration of \textbf{Algorithm} \ref{MM Algorithm} with $s=s^*$ and initialization parameter $\bsbeta_{\mathrm{ini}}$ the right support is identified at the $m$th iteration. Then there exists some constant $\zeta\in(0,1)$ and $\xi$ satisfying $\xi\sqrt{n-s}\rightarrow0\ (n\rightarrow\infty)$, such that 
\[
\Vert\bsbeta^{m+1}-\hat\bsbeta\Vert_2\leq\frac{\zeta^m}{1-\xi\sqrt{n-s}}\Vert\hat\bsbeta-\bsbeta_{\mathrm{ini}}\Vert_2+\frac{\xi\sqrt{n-s}}{1-\xi\sqrt{n-s}}s^{\frac{3}{2}}O_p\left(\frac{o((logs)^6)}{\sqrt n}\right),
\]
under the \textbf{Condition} \ref{cond.1} and \ref{cond3.2}.
\end{theorem} 

Our \textbf{Theorem} \ref{Theorem 2.} sharply contrasts \textbf{Theorem} 1 in \citep{zhang20212}, \textbf{Theorem} 1 in \citep{chen2021analysis} and their earlier counterparts, where estimation is scale with $o(n^{-\alpha})$ for some $\alpha\in(0,1)$ instead of a contraction mapping. Moreover, there is no regularization tuning parameter arisen compared with \citep{zhang20212} and no assumption of the knowledge of the active set in advance mentioned in \citep{chen2021analysis} because in section \ref{sec:selection}, we will derive methods for choosing the right active set. When $m\rightarrow\infty$, the solution $\bsbeta^m$ from \textbf{Algorithm} \ref{MM Algorithm} will fit an Erd\H{o}s-R\'enyi model accurately. This can be seen when the probability $\frac{exp(\beta_i+\beta_j)}{1+exp(\beta_i+\beta_j)}$ is restricted with $\beta_i=\beta_j=\beta_0$.

Instead of the traditional Newton method in \citep{zhang20212}, all these efforts come from our novel technical proof, which utilizes Majorization Minization(MM) algorithm via a fine-property surrogate function. In the paper \citep{balakrishnan2017statistical}, two key conditions are proposed to claim a contraction of each iteration. More details are given in \ref{sec: appendix}.

The result of \textbf{Theorem} \ref{Theorem 2.} establishes that each iteration in the \HP{} MM algorithm is a contraction mapping with the contraction coefficient $\zeta$. This result is the base of proving the convergence rate between the true value $\bsbeta^*$ and $\bsbeta^{m+1}$, which will be stated in Corollary \ref{Corollary 2.1.}.

Next we present a new $\ell_2$ distance between the algorithm solution and the truth $\bsbeta^*$.
\begin{corollary}\label{Corollary 2.1.}
Let $L:=\max_{1\leq i\leq n}\vert\beta_i^*\vert=o(loglogs)$. If the right sparsity level is chosen, i.e. $s=s^*$ and the right support is identified at the $m$th iteration, then the solution of the maximum liklelihood exists and \[ \Vert\bsbeta^{m+1}-\bsbeta^*\Vert_2\leq\frac{\zeta^m}{1-\xi\sqrt{n-s}}\Vert\hat\bsbeta-\bsbeta_{\mathrm{ini}}\Vert_2+\frac{1}{1-\xi\sqrt{n-s}}s^{\frac{3}{2}}O_p\left(\frac{o((logs)^6)}{\sqrt n}\right) , \] where $\zeta,\xi$ and $\bsbeta^{m+1}$ is defined in \textbf{Theorem} \ref{Theorem 2.}.
\end{corollary}

Without the appearance of regularization parameters, this bound of $O\left(\sqrt{\frac{s^3}{n}}\right)$ multiplied by $o((logs)^6)$ plus an order of $\zeta^m\rightarrow0$ has almost no extra assumptions but a mild order constrain on $L$ which is also mentioned in \citep{yan2013central}. This corollary considers the $\ell_2$ distance between the MLE $\hat \bsbeta$ and the truth $\bsbeta^*$ by Taylor expansion with the remainder tending to zero, under the conclusion of consistency of MLE in Mathematical Statistics. Specifically, we will write the distance between $\hat \bsbeta$ and $\bsbeta^*$ as the sum of random variable and use Central Limit Theorem to derive the convergence rate. The detailed proof will be given in \ref{sec: appendix}.

For the selection of tuning parameter $s$, we postpone the discussion in \textbf{section \ref{sec:selection}}, using the information criterion called \GICstar{}.

\subsection{Consistent tuning parameter selection with \GICstar{}}\label{sec:selection}

Before we present the main results, we first introduce some necessary notation and technical conditions.

Let $\bsbeta^*$ be the true parameter and its corresponding heterogeneous set is $\calA^* = \{j, \bsbeta^*_j \neq \beta_0^*\}$. For any parameter $\bbeta$ with heterogeneous set $\calA$, define its Kullback-Leibler (KL) divergence as
\begin{equation}\label{eq:kl-divergence}
    I(\bsbeta(\mcalA))=\sum_{1\leq i<j\leq n}\left\{p_{ij}^*(\bsbeta^*_{i}+\bsbeta^*_{j}-\bsbeta_i-\bsbeta_j)-  \log\left(\frac{1+ \exp(\bsbeta^*_{i}+\bsbeta^*_{j})}{1+\exp({\bsbeta_i+\bsbeta_j})}\right)\right\},
\end{equation}
%
where $p_{ij}^*= \exp(\bsbeta^*_{i}+\bsbeta^*_{j})/(1+\exp(\bsbeta^*_{i}+\bsbeta^*_{j}))$ is the true connecting probability for nodes $i$ and $j$.  For any pre-specified sparsity level $s$, we define an measure of the smallest gap between the true heterogeneous effect and the homogeneous effect $\beta_0^*$ by
\begin{equation}\label{eq:minimum-kl-gap}
    \delta_n(s) =\inf_{\mcalA\not\supset\mcalA_0,\vert\mcalA\vert\leq s}\frac{2}{n(n-1)}I\left(\bsbeta^{*}(\mcalA)\right),
\end{equation} 
where the $i$-the element of $\bsbeta^{*}(\mcalA)$ equals to $\bsbeta_i^*$ if $i \in \mcalA$, and equals to $\beta_0$ if $i \not\in \mcalA$.

Since the KL divergence is non-negative and $I(\bsbeta^*)=0$, the true parameter $\bsbeta^*$ is a global minimizer. With the shorthand of $I(\bsbeta(\mcalA))$, we define the KL divergence minimizer\[ \bsbeta^{KL}(\mcalA):=arg\min_{\{\mathrm{supp}(\bsbeta-\beta_0\mathbf{1})=\mcalA\}}I(\bsbeta(\mcalA)). \]

\begin{condition}\label{cond3.1}
For every $\mcalA$ satisfying $\vert\mcalA\vert\leq K$, we need $K=O(\sqrt{n})$. 
\end{condition}

Intuitively, we can view $K$ as maximum of searching range $s_{max}$ in \textbf{Algorithm} \ref{Algorithm 2}. $K$ is set as the real sparsity level $s_0=o(n^{1-\alpha})$ for some $\alpha\in[0,1)$ in \citep{chen2021analysis}. In fact, it will do harm to the whole efficiency of the algorithm itself when $K$ is too large for the time expense cost in the process of locating the smallest \GICstar{} value.

Moreover, for identifiability, $\bsbeta^{KL}(\mcalA)$ defined is the unique minimizer for all $\mcalA$ satisfying $\vert\mcalA\vert\leq K$ as \citep{fan2013tuning} pointed out some sufficient conditions for a generalized linear model with the conditional density function of $Y_i$ given $\mathbf{x}_i$\[f_i(y_i;\theta_i,\phi)=exp(y_i\theta_i-b(\theta_i)+c(y_i,\phi)):\]
\begin{itemize}
\item $\mathbf{X}_\mcalA$ has the full rank where $\mathbf{X}_\mcalA$ is the design matrix corresponding to the model $\mcalA$.
\item The function $b(\theta)$ is three times fifferentiable with $c_0\leq b''(\theta)\leq c_0^{-1}$ and $\vert b'''(\theta)\vert \leq c_0^{-1}$ in its domain for some constant $c_0>0$.
\end{itemize}

Under the \HP{} formulation of the $\beta$-model, the conditions above are easy to verify with the design matrix $\mathbf{X}=(y_1,\cdots,y_n)\in\bbR^{\frac{n(n-1)}{2}\times n}$ and $b(\theta)=\log(1+exp(\theta))$:
\begin{itemize}
\item We note that $\mathrm{rank}(\mathbf{X}^\top\mathbf{X})=\mathrm{rank}(\mathbf{X})$ and \[\mathbf{X}^\top\mathbf{X}=\begin{bmatrix}
n-1 & 1 & 1 & \cdots & 1 \\
1 & n-1 & 1 & \cdots & 1 \\
1 & 1 & n-1 &\cdots & 1 \\
\vdots & \vdots & \vdots &\ & \vdots \\
1 & 1 & 1 & \cdots & n-1 
\end{bmatrix}\in\bbR^{n\times n}.\] So that $\mathrm{det}(\mathbf{X}^\top\mathbf{X})=2(n-1)(n-2)^{n-1}\neq 0\Rightarrow\mathbf{X}$ has a full rank. The similar discussion is available to the matrix $\mathbf{X}_\mcalA$.
\item $b'''(\theta)=\frac{exp(\theta)-exp(2\theta)}{(1+exp(\theta))^3}$. It is easy to check that $b'''(\theta)\in\left[\frac{-5-3\sqrt 3}{(3+\sqrt 3)^3},\frac{3\sqrt 3-5}{(3-\sqrt 3)^3}\right]$.
\end{itemize}
Thus, it follows immediately that $\bsbeta^{KL}(\mcalA)=\bsbeta^*$ for all $\mcalA\supset\mcalA^*$ with $\vert\mcalA\vert\leq K$, and consequently $I(\bsbeta^{KL}(\mcalA))=0$.

\begin{condition}\label{cond3.2}
The real parameter $\bsbeta^*$ and $\bsbeta^{KL}(\mcalA)$ are bounded in the sense of $\ell_\infty$ norm for the ease of proof. Consequently, $\Vert\bsbeta^{KL}(\mcalA)\Vert^2_2\leq C\vert\mcalA\vert$ for some constant $C$.
\end{condition}

It is common to see the constraint put on $\bsbeta^{KL}(\mcalA)$ such as \citep{fan2013tuning}. Meanwhile, we can see that $\Vert\hat\bsbeta^{ora}(\mcalA)\Vert_\infty\leq\Vert\hat\bsbeta^{ora}(\mcalA)-\bsbeta^*\Vert_\infty+\Vert\bsbeta^*\Vert_\infty\leq \vert\mcalA\vert^{\frac{3}{2}}O_p\left(\frac{o((log\vert\mcalA\vert)^6)}{\sqrt n}\right)+C\leq \tilde C$ as $n\rightarrow\infty$ for some constant $\tilde C$ by \textbf{Corollary} \ref{Corollary 1.}. As a result, we can also obtain the conclusion $\Vert\hat\bsbeta^{ora}(\mcalA)\Vert_2^2\leq C\vert\mcalA\vert$ for some constant $C$.

\begin{condition}\label{cond:delta}
     $\frac1{\delta_n(s_{\max})} = o\left(\frac{ n(n-1)}{s_{\max} \log n \log\log n } \right)$.
\end{condition}

Next theorem guarantees that \textbf{Algorithm 2} can identify the true active with an overwhelming probability.

\begin{theorem}\label{Theorem 3.3.}
Suppose $(\hat \bsbeta,\hat \mcalA)$ is the solution of \textbf{Algorithm} \ref{Algorithm 2} with $s_{\max}$. Then under Conditions \ref{cond3.1}-\ref{cond:delta}, with probability tending to 1 and sufficiently large $n$, \textbf{Algorithm} \ref{Algorithm 2} selects the true active set, that is,  $\hat \mcalA=\mcalA^*$.
\end{theorem}

\textbf{Theorem} \ref{Theorem 3.3.} alludes that with the \GICstar{} proposed, \textbf{Algorithm} \ref{Algorithm 2} can recover the true active set, with the probability approaching 1. In contrast, almost no details were mentioned either in \citep{zhang20212} or in \citep{chen2021analysis} about how to select the regularizer $\lambda$ or the tuning parameter $s$. Whereas BIC was presented for the selection of $s$ in \citep{chen2021analysis}, yet proof for asymptotic property is lacked.

Combining \textbf{Corollary} \ref{Corollary 2.1.} and \textbf{Theorem} \ref{Theorem 3.3.}, we have the following Corollary:
\begin{corollary}\label{Corollary 1.}
Suppose $\bsbeta^{m+1}_{s_{min}}$ is the \textit{(m+1)}-th iteration after choosing $s_{min}$ via \GICstar{}, $\hat \mcalA$ is the solution of \textbf{Algorithm} \ref{Algorithm 2} and under the \textbf{Conditions} \ref{cond.1}-\ref{cond:delta}, then with probability tending to 1, the solution of the MLE $\hat \bsbeta$ exists satisfying\[ \Vert\bsbeta^{m+1}_{s_{min}}-\bsbeta^*\Vert_2\leq\frac{\zeta^m}{1-\xi\sqrt{n-s^*}}\Vert\hat\bsbeta-\bsbeta^{\mathrm{ini}}_{s_{min}}\Vert_2+\frac{1}{1-\xi\sqrt{n-s^*}}s^{*\frac{3}{2}}O_p\left(\frac{o((logs^*)^6)}{\sqrt n}\right), \]where $\bsbeta^{\mathrm{ini}}_{s_{min}}$ is the initial value chosen when the sparsity level $s_{min}$ is selected in \textbf{Algorithm} \ref{MM Algorithm} and $\zeta$ and $\xi$ are defined in \textbf{Theorem} \ref{Theorem 2.}. Moreover, we have\[ P(\hat \mcalA=\mcalA^*)\rightarrow 1, \]as $n\rightarrow\infty$. 
\end{corollary}

The detailed proof can be found in \ref{sec: GIC} and below.
\section{Simulation study}\label{sec: sim}\label{sec:simulation}

We evaluate support recovery, estimation accuracy, and computational cost. In all experiments, the true node parameters satisfy
\[
\beta_1^\star=\cdots=\beta_{s^\star}^\star=\alpha\log n,\qquad
\beta_{s^\star+1}^\star=\cdots=\beta_n^\star=\gamma\log n,
\]
with the following configurations of $(\alpha,\gamma)$:
\begin{enumerate}
\renewcommand{\labelenumi}{(\roman{enumi})}
\item $\alpha=1/5$ and $\gamma=-1/3$;
\item $\alpha=1/5$ and $\gamma=-1/2$;
\item $\alpha=1/3$ and $\gamma=-2/3$.
\end{enumerate}
We consider $n=100,150,200,250,300,350$ and four heterogeneous-set sizes:
\[
s^\star=|A^\star|\in\{2,\lfloor\sqrt{n/2}\rfloor,\lfloor\sqrt n\rfloor,\lfloor2\sqrt n\rfloor\}.
\]
Here $A^\star=\{i:\beta_i^\star\ne\beta_0^\star\}$ and $\beta_0^\star=\gamma\log n$.
The contrast $|\alpha-\gamma|$ increases from setting (i) to setting (iii). Each configuration uses $B=10$ Monte Carlo repetitions. The experiments were conducted using R and MATLAB. Complete numerical results are retained in the separate supplementary material, Tables~S1--S11.

Throughout these experiments, \HP{} denotes our Homogeneous Pursuit estimator. Our primary procedure, \HPGIC, selects the sparsity level using the proposed \GICstar; \HPBIC{} uses the same estimator with BIC. The benchmark \SBMBIC{} combines the Sparse $\beta$-Model estimation procedure of \citet{chen2021analysis} with BIC, whereas \SBMGIC{} applies our \GICstar{} to that existing estimation procedure. Thus, comparing the two criteria within a method examines sparsity selection, while comparing \HP{} and \SBM{} under the same criterion examines the estimation procedures. For estimation accuracy, we also include $\ell_2$-Newton and $\ell_2$-Gradient, the two optimization implementations of the $\ell_2$-penalized $\beta$-model estimator in \citet{zhang20212}. The runtime records identify only \HP{} and \SBM{} and do not distinguish their selection criteria.

We summarize parameter estimation using the mean absolute error $\|\widehat\beta-\beta^\star\|_1/n$, root mean squared error $\|\widehat\beta-\beta^\star\|_2/\sqrt n$, and relative error $\|\widehat\beta-\beta^\star\|_2/\|\beta^\star\|_2$. The main accuracy figure presents the reported logarithm of the relative error. To compare node parameters on the same scale, the \SBM{} estimates are transformed to the original $\beta$ parametrization as $\widehat\beta_i^{\mathrm{S}\beta\mathrm{M}}=\widehat\mu/2+\widehat\theta_i$, where $\widehat\theta_i$ is the estimated nonnegative local effect.

For the estimated heterogeneous set $\widehat A^{(b)}$ in repetition $b$, exact recovery frequency is $B^{-1}\sum_{b=1}^B\mathbf{1}\{\widehat A^{(b)}=A^\star\}$. Precision is $|\widehat A^{(b)}\cap A^\star|/|\widehat A^{(b)}|$ for a nonempty selected set, and recall is $|\widehat A^{(b)}\cap A^\star|/|A^\star|$; these measures are averaged over repetitions. Overselection and underselection frequencies are, respectively, $B^{-1}\sum_{b=1}^B\mathbf{1}\{|\widehat A^{(b)}|>|A^\star|\}$ and $B^{-1}\sum_{b=1}^B\mathbf{1}\{|\widehat A^{(b)}|<|A^\star|\}$.

\clearpage
\subsection{Simulation 1: Support recovery with \GICstar}

\begin{figure}[H]
\centering
\includegraphics[width=\linewidth]{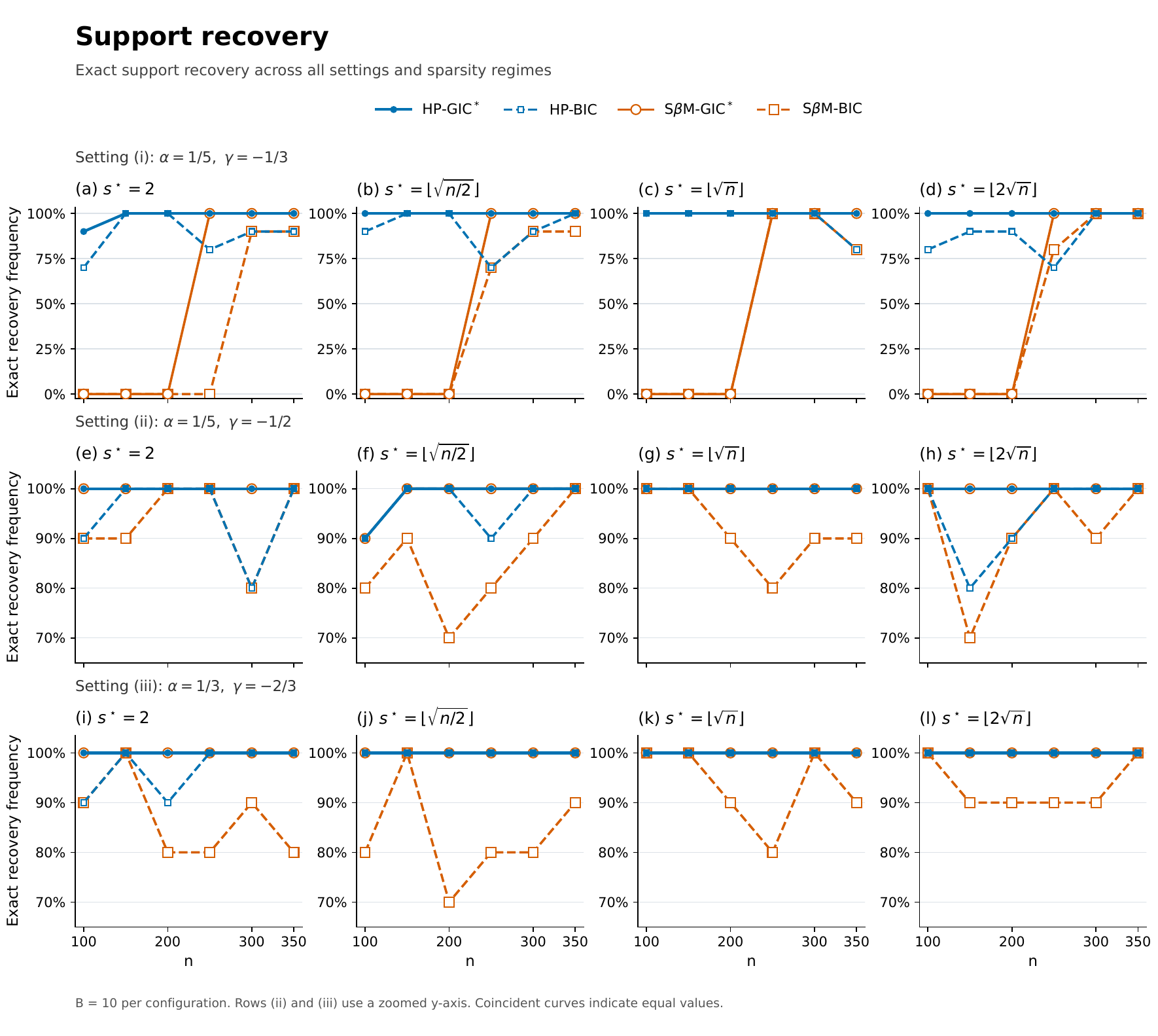}
\caption{Exact heterogeneous-set recovery over $B=10$ repetitions. Rows correspond to settings (i)--(iii), and columns to the four displayed sparsity regimes. Blue and orange distinguish \HP{} and \SBM; solid circles and dashed squares distinguish \GICstar{} and BIC. The first row shows the full frequency range; the lower two rows use a zoomed vertical range starting at $65\%$ to reveal differences near complete recovery. Coincident curves indicate equal reported frequencies. Full selection metrics appear in Supplementary Tables~S1--S3.}
\label{fig:support-recovery}
\end{figure}

Figure~\ref{fig:support-recovery} compares exact recovery frequencies for the four combinations of estimator and selection criterion.

Figure~\ref{fig:support-recovery} shows that \HPGIC{} achieves consistently high recovery frequencies across the reported configurations. Its advantage is most apparent in setting (i) for smaller networks: at $n=100$ and $s^\star=2$, the recovery frequency is $0.9$ for \HPGIC, compared with $0.7$ for \HPBIC{} and $0$ for both \SBM{} variants. In settings (ii) and (iii), \SBMGIC{} frequently matches \HPGIC, while BIC-based selection is less stable in several configurations. These comparisons support the usefulness of \GICstar{} and the strong recovery performance of \HPGIC; differences of $0.1$ represent one replication with the present $B=10$.

\clearpage
\subsection{Simulation 2: Estimation accuracy of \HP}

\begin{figure}[H]
\centering
\includegraphics[width=\linewidth,height=0.67\textheight,keepaspectratio]{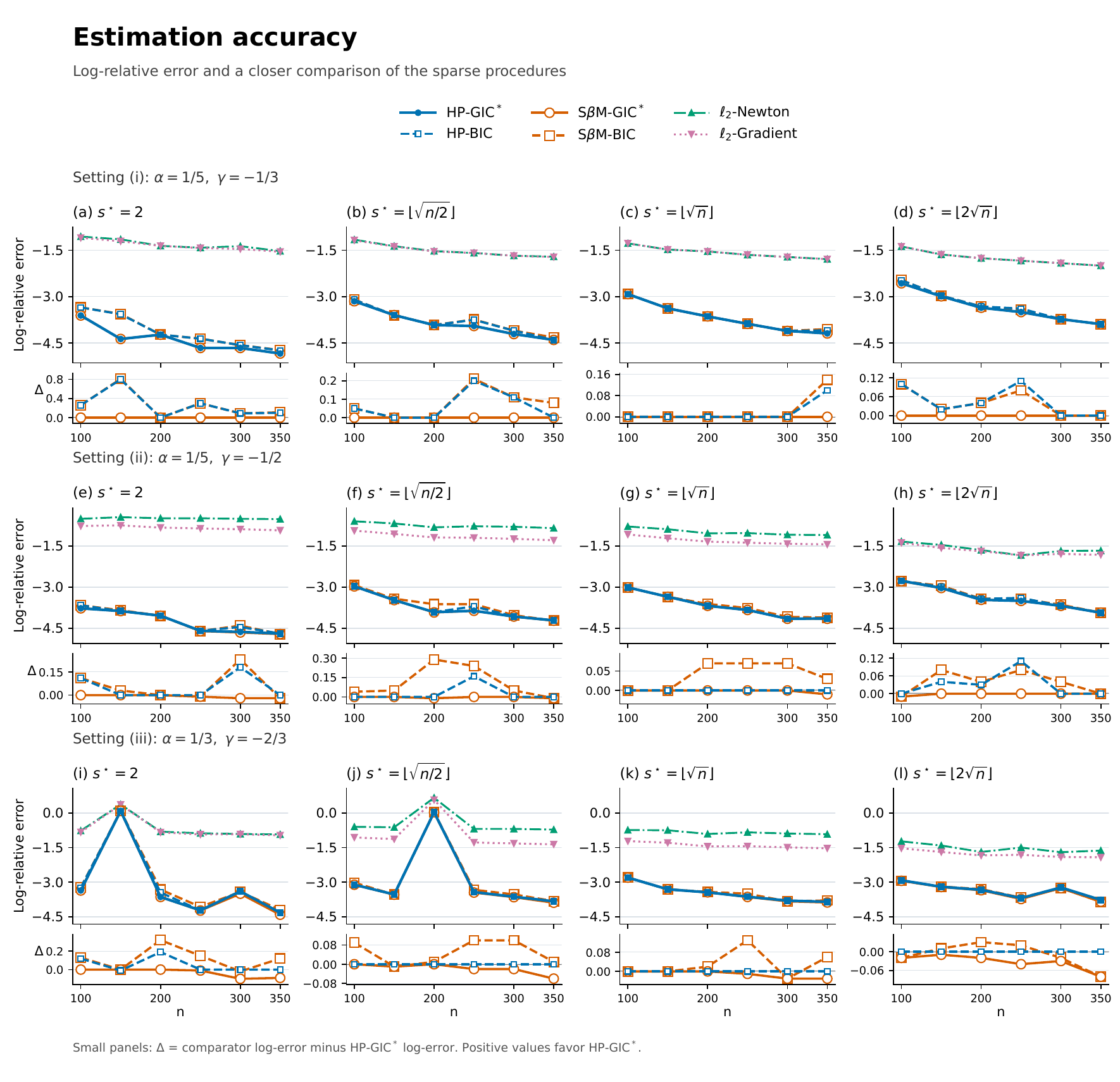}
\caption{Reported log-relative estimation error; lower values are better. Rows represent settings (i)--(iii) and columns the four sparsity regimes. The large panels compare all six procedures using a common vertical scale within each row. Each small panel shows $\Delta_m=E_m-E_{\mathrm{HP\text{-}GIC}^{*}}$ for the three other sparse procedures, where $E$ is the reported log-relative error. Positive differences favor \HPGIC; zero indicates a tie. Difference panels use their own vertical ranges. Complete error results appear in Supplementary Tables~S4--S9.}
\label{fig:estimation-accuracy}
\end{figure}

Figure~\ref{fig:estimation-accuracy} compares reported log-relative errors and displays the smaller differences among the four sparse procedures separately.

Figure~\ref{fig:estimation-accuracy} shows smaller reported errors for \HPGIC{} than for both $\ell_2$-penalized implementations across all displayed configurations. Using \GICstar{} can also improve on BIC: in setting (i), at $s^\star=2$ and $n=150$, the log-relative errors are $-4.37$ for \HPGIC{} and $-3.57$ for \HPBIC. The difference panels show comparable accuracy among the sparse procedures, with \HPGIC{} and \SBMGIC{} coinciding in $49$ of the $72$ configurations. Thus, \HPGIC{} maintains estimation accuracy comparable to the sparse benchmarks while providing the support-recovery and computational advantages illustrated in the other experiments.

\clearpage
\subsection{Simulation 3: Computational cost of \HP}

\begin{figure}[H]
\centering
\includegraphics[width=\linewidth]{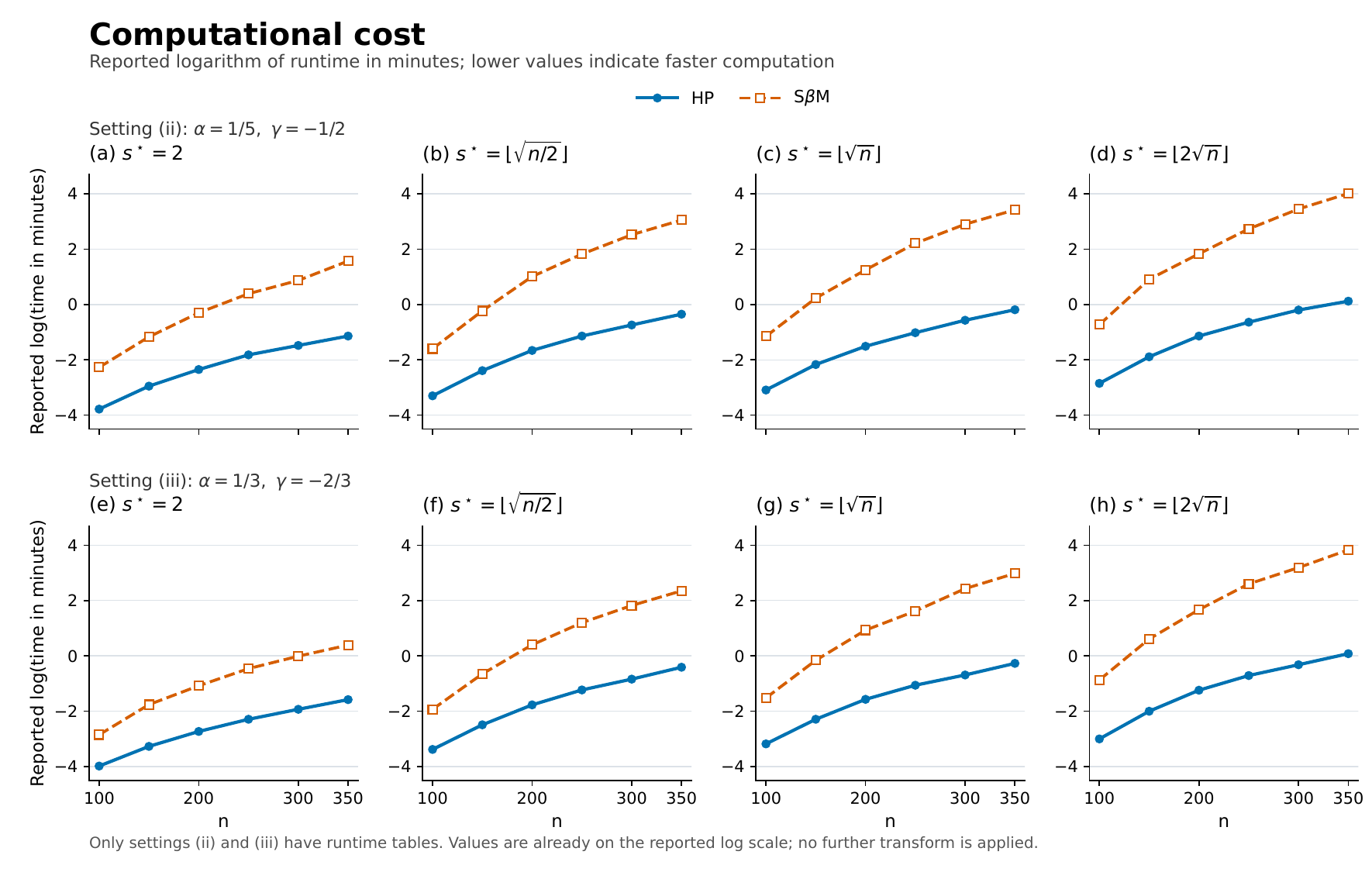}
\caption{Reported logarithm of runtime in minutes for \HP{} and \SBM. Rows represent settings (ii) and (iii), and columns the four sparsity regimes. Smaller values indicate lower computational cost. Values are plotted directly on their reported log scale. The source timings do not distinguish BIC from \GICstar{} or specify the scope of sparsity search, so this figure compares the reported estimation procedures. Complete timings appear in Supplementary Tables~S10--S11.}
\label{fig:runtime}
\end{figure}

Figure~\ref{fig:runtime} presents the available runtime comparisons. The source records contain timings for two estimation procedures, \HP{} and \SBM, in settings (ii) and (iii); they do not specify the selection criterion or whether a full sparsity search is included. We therefore retain these estimator-level labels.

Figure~\ref{fig:runtime} shows lower reported runtime for \HP{} in every displayed configuration. Computational cost increases with network size and generally with the number of heterogeneous nodes, while the gap between \HP{} and \SBM{} grows over the tested network sizes. For example, at $n=350$ and $s^\star=2$ in setting (ii), the reported log-runtime is $-1.14$ for \HP{} and $1.57$ for \SBM. These results demonstrate a computational advantage for the reported \HP{} implementation. Together with the preceding recovery and accuracy results, they support its use for homogeneous pursuit; criterion-specific end-to-end timing would be needed to quantify the additional cost of choosing sparsity by \GICstar.

\clearpage

\section{Data analysis}\label{sec:data-analysis}

In this section, we analyse the microfinance take-up example in \citet{banerjee2013diffusion} to illustrate the usefulness of \HPGIC{}. \citet{banerjee2013diffusion} investigated the role of social networks, especially the role of those pre-identified as 'leaders' (e.g. teachers, shopkeepers, savings group leaders), on households microfinancing decisions, and modelled the microfinancing decisions using a logit model.

\subsection{Data}

In 2006, data were collected for 75 rural villages in Karnataka, a state in southern India. A census of households was conducted, and a subset of individuals was asked detailed questions about the relationships they had with others in the village. This information was used to create network graphs for each village.

The social network data were collected along 12 dimensions in terms of whether individuals borrowed money from, gave advice to, helped with a decision, borrowed kerosene or rice from, lent kerosene or rice to, lent money to, obtained medical advice from, engaged socially with, were related to, went to temple with, invited to one's home, or visited another's home. A relationship between households exists if any household member indicated a relationship with members from the other household. It should be noted that no relation exists between any two households in different villages due to the nature of data collection.

In 2007, a microfinancing institution, Bharatha Swamukti Samsthe (BSS), began operations in these villages, and collected data on the households who participated in the microfinancing program. For the 43 villages that BSS entered, the total number of households is 9598 ; the average number of households for each village is 223 with a standard deviation of 56 ; and the average take-up rate for BSS is $18 \%$, with a cross-village standard deviation of $8 \%$.

For the data used in our paper, we considered the network in which two households are linked if and only if any of the 12 dimensions of social contact occurred between them. The adjacency matrix of this network is block diagonal as there exists no link between any two households in different villages. All the data sets are available through the Harvard Dataverse Network \url{https://dataverse.harvard.edu/dataset.xhtml?persistentId=hdl:1902.1/21538}.

\subsection{Household-level participation}
Following \citet{banerjee2013diffusion}, we study how social importance, in a network sense, will affect microfinance take-up decision. Using \HP, we identified heterogeneous households as those whose $\tilde \beta =\hat \beta - \hat \beta_0$ parameters are estimated as non-zero.

Since households in different villages are not connected, we allowed village-dependent $\beta_0$ parameters to capture the individual village effects in applying \HP. More precisely, for each village, we fitted the observed village network using \HPGIC, with the \SBMBIC{} procedure as the benchmark. In this data analysis, we examined $s \leq n_m / 2$ for village $m$ ( $n_m$ is the sample size of village $m$ ), that is, the maximum value of $s$ examined is half the sample size of the village. Having obtained the parameter estimates denoted as $\widehat{\beta}_m$ and $\widehat{\beta}_{0m}$ for village $m$, we used $\widehat{\beta}_{m i}$ as a covariate for household $i$ in village $m$ to model the probability that this household participated in the microfinancing program
$$
\operatorname{Parti}_{m i}=\Lambda\left(c+\theta \cdot \hat{\beta}_{m i}\right),
$$
where $\Lambda$ is the logistic function such that $\Lambda(a)=\exp(a)/(1+\exp(a))$ for $a\in\mathbb{R}$, and $c \in \mathbb{R}$ and $\theta \in \mathbb{R}$ are two unknown parameters. The role of these estimated $\beta$ 's will be referred to as $\beta$-centrality hereafter. As an alternative measure of leadership, we also examined the use of an indicator variable 'Leader', defined as $\operatorname{Leader}_{mi}:=1\left\{\tilde{\beta}_{m i}>0\right\}$. Below we suppress the dependence of $\widehat{\beta}$ and $\widehat{\beta}_{0}$ on $m$ for simplicity.

For comparison, degree centrality and eigenvector centrality, two widely used measures of the influence of a node, were also investigated. In the context of the data analysis, the degree centrality of household $i$ is $d_i$, the number of links that this household has. This is a measure of how well-connected a household is in the network. In graph theory, eigenvector centrality is a recursively defined notion of importance by associating high scores to those nodes that are connected to high-scoring nodes. Mathematically, the eigenvector centrality of the $i$ th household is the $i$ th element of $\boldsymbol{x}$, where $\boldsymbol{x}=\left(x_1, \ldots, x_n\right)^T$ is the non-negative eigenvector associated with the largest eigenvalue of the adjacency matrix, normalized to have Euclidean norm $n$. Considering these measures, we examine the six logistic specifications reported in Tables~\ref{dataana1} and~\ref{dataana1-gic}:

(1) $\operatorname{Parti}_i=\Lambda(c+\theta\widehat\beta_i)$;

(2) $\operatorname{Parti}_i=\Lambda(c+\theta(1\{\tilde\beta_i>0\}+\widehat\beta_0))$;

(3) $\operatorname{Parti}_i=\Lambda(c+\theta_1d_i+\theta_2\widehat\beta_i)$;

(4) $\operatorname{Parti}_i=\Lambda(c+\theta_1d_i+\theta_2(1\{\tilde\beta_i>0\}+\widehat\beta_0))$;

(5) $\operatorname{Parti}_i=\Lambda(c+\theta_1x_i+\theta_2\widehat\beta_i)$;

(6) $\operatorname{Parti}_i=\Lambda(c+\theta_1x_i+\theta_2(1\{\tilde\beta_i>0\}+\widehat\beta_0))$.

Here $c$, $\theta$, $\theta_1$ and $\theta_2$ are unknown coefficients. The village-dependent $\widehat\beta_0$ cannot be absorbed into the common intercept $c$. Specifications (1)--(2) examine the estimated network measures individually, while (3)--(6) assess their associations with participation after adjustment for degree or eigenvector centrality. Degree-only and eigenvector-only models are additional reference specifications and are not included in the six-column tables. All logistic regression coefficients are estimated by maximum likelihood.

\subsection{Results}
Boldface identifies \HPGIC{} results throughout the empirical tables; significance is denoted by superscripts. Tables~\ref{dataana1} and~\ref{dataana1-gic} report the benchmark and proposed estimates.

\begin{table}[H]
\centering
\begin{threeparttable}
{\small
\caption{Household participation regressions using \SBMBIC}\label{dataana1}
\begin{tabular}{lcccccc}
        \hline
		\hline		
        &(1)&(2)&(3)&(4)&(5)&(6) \\
		\hline
		\multirow{2}{*}{Degree}&&&-0.006&-0.006&& \\
        &&&(0.007)&(0.005)&& \\
        \multirow{2}{*}{Eigenvector}&&&&&$3.112^{*}$&$1.864^{.}$ \\
        &&&&&(1.191)&(1.096) \\
        \multirow{2}{*}{Beta}&$0.198^{* *}$&&$0.215^{* *}$&&$0.001$& \\
        &(0.071)&&(0.103)&&(0.048)& \\
        \multirow{2}{*}{Leader}&&$0.349^{* * *}$&&$0.420^{* * *}$&&$0.211^{.}$ \\
        &&(0.091)&&(0.126)&&(0.122) \\
		\hline
\end{tabular}
}
\begin{tablenotes}
        \footnotesize
        \item Note: Standard errors are in parentheses.
        \item ${ }^* p<0.1 \text {; }{ }^{* *} p<0.05 \text {; } { }^{* * *}p<0.01$
        \item Model setting: (1) $\operatorname{Parti}_i=\Lambda\left(c+\theta \cdot \hat{\beta}_i\right)$;\\
        (2) $\operatorname{Parti}_i=\Lambda\left(c+\theta \cdot\left(1\left\{\tilde{\beta}_i>0\right\}+\hat{\beta}_0\right)\right)$;\\
        (3) $\operatorname{Parti}_i=\Lambda\left(c+\theta_1 \cdot d_i+\theta_2 \cdot \widehat{\beta}_i\right)$;\\
        (4) $\operatorname{Parti}_i=\Lambda\left(c+\theta_1 \cdot d_i+\theta_2 \cdot\left(1\left\{\tilde{\beta}_i>0\right\}+\widehat{\beta}_0\right)\right)$;\\
        (5) $\operatorname{Parti}_i=\Lambda\left(c+\theta_1 \cdot x_i+\theta_2 \cdot \hat{\beta}_i\right)$;\\
        (6) $\operatorname{Parti}_i=\Lambda\left(c+\theta_1 \cdot x_i+\theta_2 \cdot\left(1\left\{\tilde{\beta}>0\right\}+\widehat{\beta}_0\right)\right)$;
      \end{tablenotes}
\end{threeparttable}
\end{table}

\begin{table}[H]
\centering
{\small
\caption{Household participation regressions using \HPGIC}\label{dataana1-gic}
\begin{tabular}{lcccccc}
        \hline
		\hline		
        &\proposedcell{(1)}&\proposedcell{(2)}&\proposedcell{(3)}&\proposedcell{(4)}&\proposedcell{(5)}&\proposedcell{(6)} \\
		\hline
		\multirow{2}{*}{Degree}&&&\proposedcell{-0.001}&\proposedcell{0.007}&& \\
        &&&\proposedcell{(0.006)}&\proposedcell{(0.006)}&& \\
        \multirow{2}{*}{Eigenvector}&&&&&\proposedcell{$3.336^{* *}$}&\proposedcell{$3.062^{* *}$} \\
        &&&&&\proposedcell{(0.1024)}&\proposedcell{(0.983)} \\
        \multirow{2}{*}{Beta}&\proposedcell{$0.213^{*}$}&&\proposedcell{$0.127^{}$}&&\proposedcell{$-0.047^{}$}& \\
        &\proposedcell{(0.106)}&&\proposedcell{(0.134)}&&\proposedcell{(0.048)}& \\
        \multirow{2}{*}{Leader}&&\proposedcell{$0.338^{* * *}$}&&\proposedcell{$0.205^{}$}&&\proposedcell{$0.018^{}$} \\
        &&\proposedcell{(0.062)}&&\proposedcell{(0.161)}&&\proposedcell{(0.162)} \\
		\hline
\end{tabular}
}
\end{table}

\begin{table}[H]
\centering
{\small
\caption{Household participation: AUC and F1}\label{dataana4}
\begin{tabular}{llcccccc}
        \hline
		\hline		
        evaluation&Model&(1)&(2)&(3)&(4)&(5)&(6) \\
		\hline
        \multirow{2}{*}{AUC}&\SBMBIC&0.54&0.54&0.54&0.54&0.55&0.56\\
        &\proposedcell{\HPGIC}&\proposedcell{0.55}&\proposedcell{0.55}&\proposedcell{0.54}&\proposedcell{0.55}&\proposedcell{0.55}&\proposedcell{0.55} \\
        \cline{1-8}
        \multirow{2}{*}{F1}&\SBMBIC&0.23&0.25&0.2&0.25&0.25&0.24 \\
        &\proposedcell{\HPGIC}&\proposedcell{0.25}&\proposedcell{0.26}&\proposedcell{0.26}&\proposedcell{0.26}&\proposedcell{0.27}&\proposedcell{0.27}\\
		\hline
\end{tabular}
}
\end{table}

\subsection{Village-level take-up}

We estimate regressions of the form
$$
y_r=\beta_0+\beta_1 \cdot \xi_r^L+W_r^{\prime} \delta+\epsilon_r
$$
where $y_r$ is the average village level microfinance take-up, $\xi_r^L$ is a vector of network statistics for the leaders (we introduce, separately and together, degree and eigenvector centrality), and $W_r$ is a vector of village level controls.

Though it is likely to be endogenous, we also display specifications where we introduce the centrality of the leaders who have become microfinance members:
$$
y_r=\beta_0+\beta_1 \cdot \xi_r^L+\beta_2 \cdot \xi_r^{L M}+W_r^{\prime} \delta+\epsilon_r
$$
where $\xi_r^{L M}$ is a vector of the set of leaders who became microfinance members.




\begin{table}[H]
\centering
\begin{threeparttable}
{\small
\caption{Village-level take-up regressions by leader identification method}\label{dataana21}
\begin{tabular}{lcccccc}
        \hline
		\hline		
        Covariate&\shortstack{(1)\\Original}&\shortstack{(1)\\\SBMBIC}&\proposedcell{\shortstack{(1)\\\HPGIC}}&\shortstack{(2)\\Original}&\shortstack{(2)\\\SBMBIC}&\proposedcell{\shortstack{(2)\\\HPGIC}}\\
		\hline
		\multirow{2}{*}{Degree of Leaders}&-0.002&-0.003&\proposedcell{-0.003}&-0.003&-0.005&\proposedcell{$-0.005^{.}$} \\
        &(0.003)&(0.004)&\proposedcell{(0.002)}&(0.003)&(0.004)&\proposedcell{(0.002)}\\
        Fraction of&$0.387^{***}$&$0.439^{***}$&\proposedcell{$0.448^{***}$}&$0.389^{***}$&$0.437^{***}$&\proposedcell{$0.441^{***}$} \\
        Taking Leaders&(0.087)&(0.074)&\proposedcell{(0.077)}&(0.088)&(0.073)&\proposedcell{(0.076)}\\
        Eigenvector Centrality &&&&-0.320&-0.556&\proposedcell{-0.692} \\
        of Taking Leaders&&&&(0.485)&0.426&\proposedcell{(0.440)} \\
        \multirow{2}{*}{Saving}&-0.022&-0.030&\proposedcell{-0.033}&-0.024&-0.044&\proposedcell{-0.051} \\
        &(0.110)&(0.093)&\proposedcell{(0.095)}&(0.111)&(0.093)& \proposedcell{(0.094)}\\
        \multirow{2}{*}{Fraction GM}&-0.021&-0.018&\proposedcell{-0.016}&-0.021&-0.020&\proposedcell{-0.017} \\
        &(0.030)&(0.025)&\proposedcell{(0.025)}&(0.030)&(0.025)& \proposedcell{(0.025)}\\
        \multirow{2}{*}{Constant}&0.216&0.242&\proposedcell{0.262}&0.245&0.347&\proposedcell{$-0.381^{.}$} \\
        &(0.205)&(0.201)&\proposedcell{(0.194)}&(0.211)&(0.215)&\proposedcell{(0.205)} \\
        Observations &43&43&\proposedcell{43}&43&43&\proposedcell{43}\\
        R-squared &0.387&0.528&\proposedcell{0.547}&0.394&0.549&\proposedcell{0.575}\\
		\hline
\end{tabular}
}
\begin{tablenotes}
        \footnotesize
        \item Note: Dependent variable is the microfinance participation rate of the non-leander households and report heteroskedastic robust standard errors.
      \end{tablenotes}
\end{threeparttable}
\end{table}

Tables~\ref{dataana1}--\ref{dataana21} illustrate the empirical value of \HPGIC{} at both the household and village levels. In the household regressions, the estimated node effect and the Leader measure are positively associated with participation when entered individually, although these associations weaken after adjustment for other centrality measures. The reported F1 scores are modestly higher for \HPGIC, while AUC values are comparable to those of \SBMBIC{} across specifications. The clearest advantage appears in the village-level regressions: \HPGIC{} attains $R^2=0.547$ and $0.575$ in specifications (1) and (2), compared with $0.528$ and $0.549$ for \SBMBIC{} and $0.387$ and $0.394$ for the original leader-based specifications. These results indicate improved in-sample fit when leaders are identified using \HPGIC. The fraction of participating leaders also has a positive coefficient in both \HPGIC{} specifications ($0.448$ and $0.441$), suggesting that participation among the network-identified leaders is associated with participation among non-leader households. This pattern supports the usefulness of the estimated network structure for describing microfinance take-up; the observational regressions do not establish a causal diffusion effect or an out-of-sample prediction advantage.

\section{Conclusion and Discussion}
In this paper, we developed Homogeneous Pursuit (\HP) in the $\beta$-model by allowing a subset of nodes to deviate from a common baseline parameter. We formulated the problem through an $\ell_0$-constrained likelihood approach and developed a primal-dual algorithm within the majorization-minimization framework. Under suitable conditions, we established convergence guarantees for the algorithm and derived estimation error bounds. We also proposed a generalized information criterion, \GICstar, and established its consistency for selecting the sparsity level.

The simulation studies compare both estimation procedures and sparsity-selection criteria. The reported results indicate that \HPGIC{} improves support recovery in several configurations, particularly in setting (i) at smaller network sizes, while \HP{} has lower reported computational cost than \SBM. Estimation accuracy is comparable to the sparse benchmarks, with similar errors when the procedures recover similar supports. These findings support using \HPGIC{} when identifying heterogeneous nodes and estimating their effects are both of interest.

The microfinance application illustrates how \HPGIC{} connects network structure with observed participation. The proposed procedure yields higher in-sample $R^2$ than \SBMBIC{} in both reported village-level regression specifications and modestly higher household-level F1 scores, while maintaining comparable AUC values. Participation among the identified leaders is positively associated with participation among non-leader households. These results illustrate improved village-level fit and an interpretable association between network leadership and take-up, together with household-level discrimination comparable to the benchmark. These associations should not be interpreted as causal effects.

The current framework focuses on undirected, unweighted networks and represents homogeneity through a single common baseline. This formulation provides a parsimonious description of networks containing a relatively small group of heterogeneous nodes. In applications with several distinct groups of nodes, however, a single baseline may not adequately represent the underlying structure. Extending the model to accommodate multiple homogeneous groups or observed node covariates would therefore broaden its applicability.

Several theoretical and methodological questions also remain open. Our convergence analysis relies on conditions concerning the sparsity level and support identification. Characterizing when the iterative algorithm identifies the correct support, and how this behavior depends on initialization, would strengthen the theoretical understanding of the procedure. Another direction is to develop uncertainty quantification for the selected node effects and account for estimation and selection uncertainty when these effects are used in subsequent regression analyses. These extensions would support a broader range of network applications.

\bibliographystyle{plainnat}
\bibliography{bibfileDF}

@article{wu2026benchmark,
  title={{Benchmark Radar: A Living Database and Search Engine for AI Benchmarks and Evaluation}},
  author={Wu, Koutian and Zhang, Wenjun and Zhou, Junjie and Han, Pengqian and Bao, Qiming and Shang, Ergan and Wang, Jiayu and Wang, Junkai and Xu, Wanghan and Lu, Songyuanyi and others},
  journal={arXiv preprint arXiv:2609.11115},
  year={2026}
}

@article{liu2026oscar,
  title={{OSCAR: Order-aware Scoring and Calibration for AI Rankings}},
  author={Liu, You and Liu, Yue and Lu, Quanchao and Shipilov, Nick},
  journal={arXiv e-prints},
  pages={arXiv--2609},
  year={2026}
}

@article{shang2026llm,
  title={{LLM Evaluation on Unseen Questions: Contextual Multidimensional IRT Model}},
  author={Shang, Ergan and Tang, Weijing and He, Yinqiu},
  journal={arXiv preprint arXiv:2608.22295},
  year={2026}
}

@article{phan2025humanity,
  title={Humanity's last exam},
  author={Phan, Long and Gatti, Alice and Han, Ziwen and Li, Nathaniel and Hu, Josephina and Zhang, Hugh and Zhang, Chen Bo Calvin and Shaaban, Mohamed and Ling, John and Shi, Sean and others},
  journal={arXiv preprint arXiv:2501.14249},
  year={2025}
}

@article{li2026long,
  title={Long-horizon-terminal-bench: Testing the limits of agents on long-horizon terminal tasks with dense reward-based grading},
  author={Li, Zongxia and Li, Zhongzhi and Shi, Yucheng and Wang, Ruhan and Yang, Junyao and Liu, Zhichao and Wu, Xiyang and Li, Anhao and Huang, Chengsong and Shang, Ergan and others},
  journal={arXiv preprint arXiv:2607.08964},
  year={2026}
}

@article{shang2026erase,
  title={{ERASE: EaRly bAckpropagation SchEdule for Faster Training of Modern Recommendation Systems}},
  author={Shang, Ergan and Truzzi, Flavio Sales},
  journal={arXiv preprint arXiv:2608.18469},
  year={2026}
}

@article{shang2025predicting,
  title={Predicting the unseen: a diffusion-based debiasing framework for transcriptional response prediction at single-cell resolution},
  author={Shang, Ergan and Wei, Yuting and Roeder, Kathryn},
  journal={Proceedings of the National Academy of Sciences},
  volume={122},
  number={52},
  pages={e2525268122},
  year={2025},
  publisher={National Academy of Sciences}
}

@article{zhou2026asi,
  title={{ASI-Bench: At the Dawn of Artificial Superintelligence}},
  author={Zhou, Junwei and Sun, Zhen and Li, Binyu and Zhou, Jiangyu and Pan, Yuexi and Wang, Hengyu and Ren, Honghe and Jia, Xiaohan and Zhou, Xueyang and Cao, Xiaoyu and others},
  journal={arXiv preprint arXiv:2608.17271},
  year={2026}
}

@article{shang2026inference,
  title={Inference for {B}alance in {D}ynamic {S}igned {N}etworks},
  author={Shang, Ergan and Zhang, Yuan and Tang, Weijing},
  journal={arXiv preprint arXiv:2606.08786},
  year={2026}
}

@article{zhang2025genetic,
  title={{Genetic Convergence Analysis of CRISPR Perturbations Deciphers Gene Functional Similarity}},
  author={Zhang, Tianyu and Shang, Ergan and Roeder, Kathryn},
  journal={bioRxiv},
  year={2025}
}

@article{yan2013central,
  title={A central limit theorem in the $\beta$-model for undirected random graphs with a diverging number of vertices},
  author={Yan, Ting and Xu, Jinfeng},
  journal={Biometrika},
  volume={100},
  number={2},
  pages={519--524},
  year={2013},
  publisher={Oxford University Press}
}

@article{nishii1984asymptotic,
  title={Asymptotic properties of criteria for selection of variables in multiple regression},
  author={Nishii, Ryuei},
  journal={The Annals of Statistics},
  pages={758--765},
  year={1984},
  publisher={JSTOR}
}

@article{she2019cross,
  title={On cross-validation for sparse reduced rank regression},
  author={She, Yiyuan and Tran, Hoang},
  journal={Journal of the Royal Statistical Society: Series B (Statistical Methodology)},
  volume={81},
  number={1},
  pages={145--161},
  year={2019},
  publisher={Wiley Online Library}
}

@article{zhang2010nearly,
  title={Nearly unbiased variable selection under minimax concave penalty},
  author={Zhang, Cun-Hui},
  journal={The Annals of statistics},
  volume={38},
  number={2},
  pages={894--942},
  year={2010},
  publisher={Institute of Mathematical Statistics}
}

@article{schwarz1978estimating,
  title={Estimating the dimension of a model},
  author={Schwarz, Gideon},
  journal={The annals of statistics},
  pages={461--464},
  year={1978},
  publisher={JSTOR}
}

@incollection{akaike1998information,
  title={Information theory and an extension of the maximum likelihood principle},
  author={Akaike, Hirotogu},
  booktitle={Selected papers of hirotugu akaike},
  pages={199--213},
  year={1998},
  publisher={Springer}
}

@article{mukherjee2018detection,
  title={Detection thresholds for the beta -model on sparse graphs},
  author={Mukherjee, Rajarshi and Mukherjee, Sumit and Sen, Subhabrata},
  journal={The Annals of Statistics},
  volume={46},
  number={3},
  pages={1288--1317},
  year={2018},
  publisher={Institute of Mathematical Statistics}
}

@article{kolaczyk2015question,
  title={On the question of effective sample size in network modeling: An asymptotic inquiry},
  author={Kolaczyk, Eric D and Krivitsky, Pavel N},
  journal={Statistical science: a review journal of the Institute of Mathematical Statistics},
  volume={30},
  number={2},
  pages={184},
  year={2015},
  publisher={NIH Public Access}
}

@article{krivitsky2011adjusting,
  title={Adjusting for network size and composition effects in exponential-family random graph models},
  author={Krivitsky, Pavel N and Handcock, Mark S and Morris, Martina},
  journal={Statistical methodology},
  volume={8},
  number={4},
  pages={319--339},
  year={2011},
  publisher={Elsevier}
}

@article{stein2020sparse,
  title={A Sparse beta -Model with Covariates for Networks},
  author={Stein, Stefan and Leng, Chenlei},
  journal={arXiv preprint arXiv:2010.13604},
  year={2020}
}

@article{yan2016asymptotics,
  title={Asymptotics in directed exponential random graph models with an increasing bi-degree sequence},
  author={Yan, Ting and Leng, Chenlei and Zhu, Ji},
  journal={The Annals of Statistics},
  volume={44},
  number={1},
  pages={31--57},
  year={2016},
  publisher={Institute of Mathematical Statistics}
}

@article{yan2019statistical,
  title={Statistical inference in a directed network model with covariates},
  author={Yan, Ting and Jiang, Binyan and Fienberg, Stephen E and Leng, Chenlei},
  journal={Journal of the American Statistical Association},
  volume={114},
  number={526},
  pages={857--868},
  year={2019},
  publisher={Taylor \& Francis}
}

@article{zhang20212,
  title={L-2 Regularized maximum likelihood for beta -model in large and sparse networks},
  author={Zhang, Y and Wang, Q and Yan, T and Luo, J},
  year={2021}
}

@article{mcpherson2001birds,
  title={Birds of a feather: Homophily in social networks},
  author={McPherson, Miller and Smith-Lovin, Lynn and Cook, James M},
  journal={Annual review of sociology},
  pages={415--444},
  year={2001},
  publisher={JSTOR}
}

@article{barabasi2003scale,
  title={Scale-free networks},
  author={Barab{\'a}si, Albert-L{\'a}szl{\'o} and Bonabeau, Eric},
  journal={Scientific american},
  volume={288},
  number={5},
  pages={60--69},
  year={2003},
  publisher={JSTOR}
}

@article{gilbert1959random,
  title={Random graphs},
  author={Gilbert, Edgar N},
  journal={The Annals of Mathematical Statistics},
  volume={30},
  number={4},
  pages={1141--1144},
  year={1959},
  publisher={JSTOR}
}

@article{erdos1960evolution,
  title={On the evolution of random graphs},
  author={Erdos, Paul and R{\'e}nyi, Alfr{\'e}d and others},
  journal={Publ. Math. Inst. Hung. Acad. Sci},
  volume={5},
  number={1},
  pages={17--60},
  year={1960},
  publisher={Citeseer}
}

@article{erdos59a,
  author = {Erd\"{o}s, P. and R\'{e}nyi, A.},
  journal = {Publicationes Mathematicae Debrecen},
  pages = 290,
  title = {On Random Graphs I},
  volume = 6,
  year = 1959
}

@inproceedings{adamic2005political,
  title={The political blogosphere and the 2004 US election: divided they blog},
  author={Adamic, Lada A and Glance, Natalie},
  booktitle={Proceedings of the 3rd international workshop on Link discovery},
  pages={36--43},
  year={2005}
}

@techreport{pakes2017advances,
  title={Advances in Economics and Econometrics},
  author={Pakes, Ariel and Piazzesi, Monika and Samuelson, Larry and others},
  year={2017},
  institution={Cambridge University Press}
}

@article{britton2006generating,
  title={Generating simple random graphs with prescribed degree distribution},
  author={Britton, Tom and Deijfen, Maria and Martin-L{\"o}f, Anders},
  journal={Journal of statistical physics},
  volume={124},
  number={6},
  pages={1377--1397},
  year={2006},
  publisher={Springer}
}

@article{barabasi2013network,
  title={Network science},
  author={Barab{\'a}si, Albert-L{\'a}szl{\'o}},
  journal={Philosophical Transactions of the Royal Society A: Mathematical, Physical and Engineering Sciences},
  volume={371},
  number={1987},
  pages={20120375},
  year={2013},
  publisher={The Royal Society Publishing}
}

@article{chatterjee2011random,
  title={Random graphs with a given degree sequence},
  author={Chatterjee, Sourav and Diaconis, Persi and Sly, Allan},
  journal={The Annals of Applied Probability},
  volume={21},
  number={4},
  pages={1400--1435},
  year={2011},
  publisher={Institute of Mathematical Statistics}
}

@book{buhlmann2011statistics,
  title={Statistics for high-dimensional data: methods, theory and applications},
  author={B{\"u}hlmann, Peter and Van De Geer, Sara},
  year={2011},
  publisher={Springer Science \& Business Media}
}

@article{fan2013tuning,
  title={Tuning parameter selection in high dimensional penalized likelihood},
  author={Fan, Yingying and Tang, Cheng Yong},
  journal={Journal of the Royal Statistical Society: Series B (Statistical Methodology)},
  volume={75},
  number={3},
  pages={531--552},
  year={2013},
  publisher={Wiley Online Library}
}

@article{chen2021analysis,
  title={Analysis of networks via the sparse $\beta$-model},
  author={Chen, Mingli and Kato, Kengo and Leng, Chenlei},
  journal={Journal of the Royal Statistical Society: Series B (Statistical Methodology)},
  year={2021},
  publisher={Wiley Online Library}
}

@article{balakrishnan2017statistical,
  title={{Statistical guarantees for the EM algorithm: From population to sample-based analysis}},
  author={Balakrishnan, Sivaraman and Wainwright, Martin J and Yu, Bin},
  journal={The Annals of Statistics},
  volume={45},
  number={1},
  pages={77--120},
  year={2017},
  publisher={Institute of Mathematical Statistics}
}

@book{lange2016mm,
  title={MM optimization algorithms},
  author={Lange, Kenneth},
  year={2016},
  publisher={SIAM}
}

@article{wen2020bess,
  title={BeSS: an R package for best subset selection in linear, logistic and Cox proportional hazards models},
  author={Wen, Canhong and Zhang, Aijun and Quan, Shijie and Wang, Xueqin},
  journal={Journal of Statistical Software},
  volume={94},
  pages={1--24},
  year={2020}
}

@article{banerjee2013diffusion,
  title={The diffusion of microfinance},
  author={Banerjee, Abhijit and Chandrasekhar, Arun G and Duflo, Esther and Jackson, Matthew O},
  journal={Science},
  volume={341},
  number={6144},
  pages={1236498},
  year={2013},
  publisher={American Association for the Advancement of Science}
}

\clearpage
\appendix
\section{Proofs of main results}\label{sec: appendix}

\subsection{Proof of Theorem \ref{Theorem 2.}}

We make the proof in two steps, by defining the following terms:

$$\hat \bsbeta=arg\min_{\bsbeta}\ell_n(\bsbeta) \ \mathrm{such} \ \mathrm{that} \ \Vert\bsbeta-\mathbf{1}\beta_0\Vert_0=s,$$ which is the MLE, $$\hat \bsbeta^m=arg\min_{\bsbeta}S(\bsbeta|\bsbeta^{m-1}) \ \mathrm{such} \ \mathrm{that} \ \Vert\bsbeta-\mathbf{1}\beta_0\Vert_0=s,$$ and $$\bsbeta^m=arg\min_{\bsbeta}S(\bsbeta|\bsbeta^{m-1}) \ \mathrm{such} \ \mathrm{that} \ \Vert\bsbeta-\mathbf{1}\beta_0^{m-1}\Vert_0=s.$$

Here we make the reference of the conclusion of \citep{balakrishnan2017statistical} to make the estimation of $\Vert\hat \bsbeta-\hat \bsbeta^m\Vert_2$, so that we only need to verify with the definition $M(\bsbeta)=arg\min_{\bsbeta'}S(\bsbeta'|\bsbeta)$ the following two conditions:

\begin{itemize}
\item $\left\langle\nabla S(M(\bsbeta)|\hat{\bsbeta})-\nabla S(\hat \bsbeta|\hat \bsbeta),M(\bsbeta)-\hat \bsbeta\right\rangle\geq\lambda\left\Vert\hat \bsbeta-M(\bsbeta)\right\Vert^2_2,$
\item $\left\Vert\nabla S(M(\bsbeta)|\hat \bsbeta)-\nabla S(M(\bsbeta)|\bsbeta)\right\Vert_2\leq\gamma\left\Vert\bsbeta-\hat \bsbeta\right\Vert_2,$
\end{itemize}
where we require that $0<\gamma<\lambda$, so that the contraction coefficient $\zeta=\frac{\gamma}{\lambda}<1$.

It is easy to find $\left\langle\nabla S(M(\bsbeta)|\hat{\bsbeta})-\nabla S(\hat \bsbeta|\hat \bsbeta),M(\bsbeta)-\hat \bsbeta\right\rangle=\frac{n-1}{2}\Vert\hat{\bsbeta}-M(\bsbeta)\Vert^2_2$, so that $\lambda=\frac{n-1}{2}$.

For the second condition, we have $\nabla S(M(\bsbeta)|\hat \bsbeta)-\nabla S(M(\bsbeta)|\bsbeta)=\hat G-G+\frac{n-1}{2}(\bsbeta-\hat{\bsbeta})$, where $G_i=\sum_{j\neq i}\frac{exp(\beta_i+\beta_j)}{1+exp(\beta_i+\beta_j)}$ denotes the ith coordinate of $\nabla_{\bsbeta}\ell_n(\bsbeta)$. We make estimations about the following term 
\begin{flushleft}
$\sum_{i=1}^n\left((\hat G_i-G_i)+\frac{n-1}{2}(\beta_i-\hat \beta_i)\right)^2$
\end{flushleft}
\begin{align*}
=&\sum_{i=1}^n\frac{1}{4}(n-1)^2(\beta_i-\hat \beta_i)^2+\sum_{i=1}^n(n-1)(\beta_i-\hat \beta_i)\sum_{j\neq i}\left(\frac{exp(\hat \beta_i+\hat{\beta_j})}{1+exp(\hat \beta_i+\hat \beta_j)}-\frac{exp(\beta_i+\beta_j)}{1+exp(\beta_i+\beta_j)}\right) \\
+&\sum_{i=1}^n\left[\sum_{j\neq i}\left(\frac{exp(\hat\beta_i+\hat \beta_j)}{1+exp(\hat \beta_i+\hat \beta_j)}-\frac{exp(\beta_i+\beta_j)}{1+exp(\beta_i+\beta_j)}\right)\right]^2    \\
:= &\sum_{i=1}^n\frac{1}{4}(n-1)^2(\beta_i-\hat \beta_i)^2+I+II,
\end{align*}where\[ I=\sum_{i=1}^n(n-1)(\beta_i-\hat \beta_i)\sum_{j\neq i}\left(\frac{exp(\hat \beta_i+\hat{\beta_j})}{1+exp(\hat \beta_i+\hat \beta_j)}-\frac{exp(\beta_i+\beta_j)}{1+exp(\beta_i+\beta_j)}\right), \] and\[ II=\sum_{i=1}^n\left[\sum_{j\neq i}\left(\frac{exp(\hat\beta_i+\hat \beta_j)}{1+exp(\hat \beta_i+\hat \beta_j)}-\frac{exp(\beta_i+\beta_j)}{1+exp(\beta_i+\beta_j)}\right)\right]^2. \]
Now we estimate terms $I,II$ using Lagrange mean value theorem on $\frac{exp(\hat \beta_i+\hat{\beta_j})}{1+exp(\hat \beta_i+\hat \beta_j)}-\frac{exp(\beta_i+\beta_j)}{1+exp(\beta_i+\beta_j)}=C_{ij}(\hat \beta_i-\beta_i+\hat \beta_j-\beta_j)$ where $C_{ij}\in\mathrm{Image}\left(\frac{e^{x}}{(1+e^{x})^2}\right)=(0,\frac{1}{4}]$.
\begin{align*}
I&\leq(n-1)\sum_{i=1}^n\left(\beta_i-\hat \beta_i\right)\sum_{j\neq i}C_1\left[(\hat \beta_i-\beta_i)+(\hat \beta_j-\beta_j)\right] \\
&= C_1(n-1)\sum_{i=1}^n\left(\beta_i-\hat \beta_i\right)\left[(n-1)(\hat \beta_i-\beta_i)+\sum_{j\neq i}(\hat \beta_j-\beta_j)\right] \\
&= -C_1(n-1)^2\sum_{i=1}^n\left(\beta_i-\hat \beta_i\right)^2+C_1(n-1)\sum_{i=1}^n\sum_{j\neq i}\left(\beta_i-\hat \beta_i\right)\left(\hat \beta_j-\beta_j\right),
\end{align*}
and
\begin{align*}
II&\leq\sum_{i=1}^n\left[\sum_{j\neq i}C_2\left(\hat \beta_i-\beta_i+\hat \beta_j-\beta_j\right)\right]^2 \\
&=C_2^2\sum_{i=1}^n\left[\left(n-1\right)\left(\hat \beta_i-\beta_i\right)+\sum_{j\neq i}\left(\hat \beta_j-\beta_j\right)\right]^2 \\
&=C_2^2\sum_{i=1}^n\left[(n-1)^2(\hat \beta_i-\beta_i)^2+\left(\sum_{j\neq i}\left(\hat \beta_j-\beta_j\right)\right)^2+2(n-1)(\hat \beta_i-\beta_i)\sum_{j\neq i}\left(\hat \beta_j-\beta_j\right)\right] \\
&=C_2^2(n-1)^2\sum_{i=1}^n\left(\hat \beta_i-\beta_i\right)^2+2C_2^2(n-1)\sum_{i=1}^n\sum_{j\neq i}\left(\hat \beta_i-\beta_i\right)\left(\hat\beta_j-\beta_j\right)+C_2^2\sum_{i=1}^n\left(\sum_{j\neq i}\left(\hat \beta_j-\beta_j\right)\right)^2,
\end{align*}
where constant $C_1,C_2\in[a_1,a_2]\subset(0,\frac{1}{4}]$ are chosen carefully with respect to the universal bound of $C_{ij}$ to make inequations hold.

As a result, we have
\begin{flushleft}
$\sum_{i=1}^n\left((\hat G_i-G_i)+\frac{n-1}{2}(\beta_i-\hat \beta_i)\right)^2$
\end{flushleft}
\begin{align*}
&\leq\left(\frac{1}{4}-C_1+C_2^2\right)\left(n-1\right)^2\sum_{i=1}^n\left(\beta_i-\hat \beta_i\right)^2+(2C_2^2-C_1)(n-1)\sum_{i=1}^n\sum_{j\neq i}\left(\hat \beta_i-\beta_i\right)\left(\hat \beta_j-\beta_j\right) \\
&+C_2^2\sum_{i=1}^n\left(\sum_{j\neq i}\left(\hat \beta_j-\beta_j\right)\right)^2. \\
\end{align*}
In order to make the estimation of $\gamma$ small enough, we choose $C_1=a_1$ and $C_2=a_2$ hoping that $2a_2^2<a_1\leq a_2\Rightarrow a_2<\frac{1}{2}$ which is statisfied because of $a_1,a_2\in(0,\frac{1}{4}]$.

Moreover, using Cauchy-Schwarze inequality, we obtain 
\begin{align*}
C_2^2\sum_{i=1}^n\left(\sum_{j\neq i}\left(\hat \beta_j-\beta_j\right)\right)^2&\leq a_2^2\sum_{i=1}^n\sum_{j\neq i}(n-1)(\hat \beta_j-\beta_j)^2\\
&=a_2^2(n-1)^2\sum_{i=1}^n(\hat\beta_i-\beta_i)^2
.
\end{align*}

Now we have derive that 
\begin{flushleft}
$\sum_{i=1}^n\left((\hat G_i-G_i)+\frac{n-1}{2}(\beta_i-\hat \beta_i)\right)^2$
\end{flushleft}
\begin{align*}
&\leq\left(\frac{1}{4}-a_1+2a_2^2\right)(n-1)^2\sum_{i=1}^n\left(\beta_i-\hat \beta_i\right)^2+(2a_2^2-a_1)(n-1)\sum_{i=1}^n\sum_{j\neq i}\left(\hat \beta_i-\beta_i\right)\left(\hat \beta_j-\beta_j\right) \\
&\leq\left(\frac{1}{4}-a_1+2a_2^2\right)(n-1)^2\sum_{i=1}^n\left(\beta_i-\hat \beta_i\right)^2+\left(\frac{-2a_2^2+a_1}{2}\right)(n-1)\sum_{i=1}^n\sum_{j\neq i}\left((\hat \beta_i-\beta_i)^2+(\hat\beta_j-\beta_j)^2\right)
\end{align*}
via $\vert ab\vert\leq\frac{a^2+b^2}{2}$. Observing that \begin{align*}
\sum_{i=1}^n\sum_{j\neq i}\left((\hat \beta_i-\beta_i)^2+(\hat\beta_j-\beta_j)^2\right)&=\sum_{i=1}^n\left((n-1)(\hat \beta_i-\beta_i)^2+\sum_{j\neq i}(\hat \beta_j-\beta_j)^2\right) \\
&=2(n-1)\sum_{i=1}^n\left(\hat \beta_i-\beta_i\right)^2.
\end{align*}
Finally, we conclude $$\sum_{i=1}^n\left((\hat G_i-G_i)+\frac{n-1}{2}(\beta_i-\hat \beta_i)\right)^2\leq\frac{1}{4}(n-1)^2\sum_{i=1}^n\left(\hat \beta_i-\beta_i\right)^2.$$

All the inequalities cannot attain equal at the same time, so that $\gamma=\sigma\cdot\frac{n-1}{2}$ with $\sigma\in(0,1)$ and $\zeta=\frac{\gamma}{\lambda}=\sigma<1$, so that by \citep{balakrishnan2017statistical} we have \[\Vert\hat \bsbeta-\hat \bsbeta^{m+1}\Vert_2\leq\zeta^m\Vert\hat \bsbeta-\bsbeta_{\mathrm{ini}}\Vert_2.\tag{A.1.0}\label{A.1.0}\]

By the process of the Algorithm \ref{MM Algorithm} and assuming that the active set has been identified at the $m$th iteration, we conclude that $\Vert\hat \bsbeta^{m+1}-\bsbeta^{m+1}\Vert_2^2=(n-s)(\beta_0^{m}-\beta_0)^2$, so that we have \[\Vert\bsbeta^{m+1}-\hat \bsbeta\Vert_2\leq\zeta^m\Vert\bsbeta_{\mathrm{ini}}-\hat \bsbeta\Vert_2+\sqrt{n-s}\vert\beta_0^m-\beta_0\vert\tag{A.1.1}\label{A.1.1}.\] 
Also, without loss of generality, we assume that $\mcalA^{m+1}=\mcalA^*=[s]$. According to \textbf{Algorithm} \ref{MM Algorithm}, we have $\beta_0^m-\beta_0=f(\bsbeta^{m+1}_{\mcalA^{m+1}})-f(\bsbeta^*_{\mcalA^*})$, where $f(\bsbeta)=\frac{1}{2}log\frac{p}{1-p}=\frac{1}{logit(p)}$ with  \[p=p(\bsbeta)=\frac{\sum_{i=1}^nd_i-2\sum_{i=1}^sd_i+\sum_{i\neq j,i,j=1}^s\frac{exp(\beta_i+\beta_j)}{1+exp(\beta_i+\beta_j)}}{(n-s)(n-s-1)}.\] Obviously, $\frac{\partial p}{\partial\beta_i}=\frac{1}{(n-s)(n-s-1)}\sum_{j\neq i}^s\frac{exp(\beta_i+\beta_j)}{(1+exp(\beta_i+\beta_j))^2}$, so that \[\left\vert\frac{\partial p}{\partial\beta_i}\right\vert\leq\frac{s-1}{4(n-s)(n-s-1)}\tag{A.1.2}\label{A.1.2}\]by mean inequality. By chain rule, we have $\nabla f=\frac{1}{2p(1-p)}\nabla p$. Consequently, 
\begin{align*}
\Vert\nabla f\Vert_2&=\frac{1}{2p(1-p)}\Vert\nabla p\Vert_2\leq\frac{1}{2p(1-p)}\sqrt s\sqrt{\frac{s-1}{4(n-s)(n-s-1)}} \\
&=\frac{1}{4p(1-p)}\sqrt{\frac{s(s-1)}{(n-s)(n-s-1)}}:=\xi.\tag{A.1.3}\label{A.1.3}
\end{align*}where the inequality is the result of \ref{A.1.2}.

Finally, by Lagrange Mean Value Theorem, we have
\begin{align*}
\left\Vert\hat\bsbeta^{m+1}-\bsbeta^{m+1}\right\Vert_2&\leq\sqrt{n-s}\cdot\xi\cdot\Vert\bsbeta^{m+1}_{\mcalA^{m+1}}-\bsbeta^*_{\mcalA^*}\Vert_2\\
&\leq\xi\sqrt{n-s}\left(\Vert\bsbeta^{m+1}-\hat\bsbeta^{m+1}\Vert_2+\Vert\hat\bsbeta^{m+1}-\hat\bsbeta\Vert_2+\Vert\hat\bsbeta-\bsbeta^*\Vert_2\right),\\
\tag{A.1.4}\label{A.1.4}
\end{align*}where the final inequality is by triangle inequality.

Now recall $\Vert\hat \bsbeta-\hat \bsbeta^{m+1}\Vert_2\leq\zeta^m\Vert\hat \bsbeta-\bsbeta_{\mathrm{ini}}\Vert_2$ by \ref{A.1.0} and $\Vert\hat\bsbeta-\bsbeta^*\Vert_2\leq s^{\frac{3}{2}}O_p\left(\frac{o((logs)^6}{\sqrt n}\right)$ by \textbf{Corollary} \ref{Corollary 2.1.}. By plugging them into \ref{A.1.4} and re-arranging terms, we get\[\left\Vert\hat\bsbeta^{m+1}-\bsbeta^{m+1}\right\Vert_2\leq\frac{\xi\sqrt{n-s}}{1-\xi\sqrt{n-s}}\left(\zeta^m\Vert\hat\bsbeta-\bsbeta_{\mathrm{ini}}\Vert_2+s^{\frac{3}{2}}O_p\left(\frac{o((logs)^6)}{\sqrt n}\right)\right).\]

Then plugging the term above into \ref{A.1.1} gets what we claimed in \textbf{Theorem} \ref{Theorem 2.}.

What is left now is to verify the order of $\xi$ which is defined in \ref{A.1.3}. By the definition of $\xi=\frac{1}{4p(1-p)}\sqrt{\frac{s(s-1)}{(n-s)(n-s-1)}}$, we only need to prove that $\xi\sqrt{n-s}\rightarrow0$ as $n\rightarrow\infty$.

 Under \textbf{Condition} \ref{cond3.2}, we can let $\bsbeta$ in a sufficiently large closed ball, which is a compact set. Then we can let $\frac{exp(\beta_i+\beta_j)}{1+exp(\beta_i+\beta_j)}\geq t$ and the lower bound can be obtained on the bounbdary of the compact set. 

For one thing, we can see that\[p\leq\frac{\sum_{i=1}^nd_i+\sum_{i\neq j,i,j=1}^s\frac{exp(\beta_i+\beta_j)}{1+exp(\beta_i+\beta_j)}}{(n-s)(n-s-1)}\leq\frac{O(n^{2-\rho})+s^2}{(n-s)(n-s-1)}<1\tag{A.1.5}\label{A.1.5}\] as $n\rightarrow\infty$, where the first inequality is the result of \textbf{Condition} \ref{cond.1}, meaning $1-p$ cannnot create 0 in the denominator.

Then, we have \begin{align*}
\xi\sqrt{n-s}&=\frac{1}{p}\sqrt{\frac{s(s-1)}{(n-s)(n-s-1)}}\\
&=\frac{(n-s)(n-s-1)}{\sum_{i=1}^nd_i-2\sum_{i=1}^sd_i+\sum_{i\neq j,i,j=1}^s\frac{exp(\beta_i+\beta_j)}{1+exp(\beta_i+\beta_j)}}\sqrt{\frac{s(s-1)}{n-s-1}}\\
&\leq\frac{O(sn^{\frac{3}{2}})}{ts^2+cn^{\frac{3}{2}+c'}}\rightarrow0 \ \mathrm{as}\ n\rightarrow\infty,\tag{A.1.6}\label{A.1.6}
\end{align*}
where \textbf{Condition} \ref{cond.1} results in the last inequality.

Combining \ref{A.1.5} with \ref{A.1.6} finishes the final part of the proof. 

\subsection{Proof of Corollary \ref{Corollary 2.1.}}
By the definition of the MLE, we have for $i=1,\cdots,s$,$$d_i=\sum_{j=1,j\neq i}^s\frac{exp(\hat \beta_i+\hat \beta_j)}{1+exp(\hat \beta_i+\hat \beta_j)}.$$ We define $m_{ij}=\frac{exp(\beta^*_i+\beta^*_j)}{(1+exp(\beta^*_i+\beta^*_j))^2}$ for $i,j\in[s]$ and $m_{ij}:=m_{i0}=\frac{exp(\beta^*_i+\beta_0)}{(1+exp(\beta^*_i+\beta_0))^2}$ for $j\notin[s]$ as the covariance matrix of $d=(d_1,\cdots,d_n)^\top$. Particularly, we have $m_{ii}=\sum_{j\neq i,j=1}^n$ for $i\in[s]$. Then by the equations above and Taylor expansion, we have \[ d_i-\bbE d_i=\sum_{j=1,j\neq i}^s(m_{ij}(\hat \beta_i+\hat \beta_j-\beta^*_i-\beta^*_j)+h_{ij})+(n-s)(\hat \beta_i-\beta^*_i)m_{i0}+(n-s)h_{i0},\]where $h_{ij}=\frac{e^{\beta^*_i+\beta^*_j+\theta_{ij}\hat \gamma_{ij}}(1-e^{\beta^*_i+\beta^*_j+\theta_{ij}\hat \gamma_{ij}})}{2(1+e^{\beta^*_i+\beta^*_j+\theta_{ij}\hat \gamma_{ij}})^3}\hat\gamma_{ij}^2$, $h_{i0}=\frac{e^{\beta^*_i+\beta_0+\theta_i\hat \gamma_i}(1-e^{\beta^*_i+\beta_0+\theta_i\hat \gamma_i})}{2(1+e^{\beta^*_i+\beta_0+\theta_i\hat \gamma_i})^3}(\hat \beta_i-\beta^*_i)^2$ with $\hat \gamma_{ij}=\hat \beta_i+\hat \beta_j-\beta^*_i-\beta^*_j, \hat \gamma_i=\hat \beta_i-\beta^*_i$. Therefore, we have \[ d-\bbE d=M_s(\hat \beta_S-\beta_S)+h,\] where $M_s\in\bbR^{s\times s}$ is a submatrix of $M=(m_{ij})_{i,j=1}^n$ (thus non-singular) and $h=(h_1,\cdots,h_s)^\top$ with $h_i=\sum_{j=1,j\neq i}^sh_{ij}+(n-s)h_{i0}$. So far, we derive that \[ \hat \bsbeta_S-\bsbeta^*_S=M_s^{-1}(d-\bbE d)+M_s^{-1}h\tag{A.2.1}\label{A.2.1},\]where $\bsbeta^*_S=(\beta_1^*,\cdots,\beta_s^*)^\top, \hat\bsbeta=(\hat \beta_i,\cdots,\hat \beta_s)^\top$, and then we make upper bound for the two parts above to obtain the convergence rate.

By Central Limit Theorem (CLT), we have $m_{ii}^{-\frac{1}{2}}(d_i-\bbE d_i)\overset{d}{\rightarrow}\mcalN(0,1)=O_p(1)$. Then \[\Vert d-\bbE d\Vert_2\leq\sqrt s O_p\left(\max_{1\leq i\leq s}\sqrt{m_{ii}}\right)\tag{A.2.2}\label{A.2.2}.\]Moreover, in order to approximate $M_s^{-1}$ well, we define the matrix $Q=(q_{ij})_{i,j=1}^n$ with $q_{ij}=\frac{\delta_{ij}}{m_{ii}}-\frac{1}{m_{\cdot\cdot}}$ with $m_{\cdot\cdot}=\sum_{i,j=1, i\neq j}^n m_{ij}.$ We use $Q_s\in\bbR^{s\times s}$ to make a good approximation of $M_s$. By the reference of \citep{yan2013central}, we have \[\Vert M_s^{-1}-Q_s\Vert_\infty\leq O\left(\frac{e^{6L}}{(n-1)^2}\right),\] where $\Vert A\Vert_\infty=\max_{i,j}\vert a_{ij}\vert$ for a matrix $A=(a_{ij})$. Observing that $\frac{1}{\sqrt n}\Vert A \Vert_\infty\leq\Vert A\Vert_{op}\leq n\Vert A \Vert_\infty$, we obtain \[\Vert M_s^{-1}\Vert_{op}\leq\Vert M_s^{-1}-Q_s\Vert_{op}+\Vert Q_s\Vert_{op}\leq sO\left(\frac{e^{6L}}{(n-1)^2}\right)+s\max_{i,j}\left\vert\frac{\delta_{ij}}{m_{ii}}-\frac{1}{m_{\cdot\cdot}}\right\vert\tag{A.2.3}\label{A.2.3}.\]

Furthermore, we claim that $$M_s^{-1}h=o_p(1).$$ By observing that $\vert\frac{e^x(1-e^x)}{2(1+e^x)^3}\vert\leq\frac{1}{2}$, we have $\vert h_{ij}\vert\leq\lambda_s^2$ and $\vert h_{i0}\vert\leq\frac{1}{2}\lambda_s^2$ with the notation $\lambda_s=\max_{1\leq i\leq s}\vert \hat \beta_i-\beta^*_i\vert.$ So that \[\vert h_i\vert\leq\left(\frac{n+s}{2}-1\right)\lambda_s^2.\] Consider $(M_s^{-1}h)_i=((M_s^{-1}-Q_s+Q_s)h)_i:=(F_sh)_i+(Q_sh)_i$ where $F_s=M_s^{-1}-Q_s$. For one thing, \[\vert(F_sh)_i\vert\leq\Vert F_s\Vert_\infty\cdot(s\max_i\vert h_i\vert)\leq O\left(\frac{e^{6L}}{(n-1)^2}s\left(\frac{n+s}{2}-1\right)\lambda_s^2\right)=\lambda_s^2O\left(\frac{s\left(\frac{n+s}{2}-1\right)}{(n-1)^2}e^{6L}\right)\tag{A.2.4}\label{A.2.4}.\] For another, $(Q_sh)_i=\sum_{j=1}^sq_{ij}h_j=\frac{1}{m_{ii}}h_i-m_{\cdot\cdot}^{-1}\sum_{j=1}^sh_j$. Noting that $$m_{ii}\in\left[\frac{(n-1)e^{2L}}{(1+e^{2L})^2},\frac{n-1}{4}\right], \ m_{\cdot\cdot}\geq\frac{n(n-1)e^{2L}}{(1+e^{2L})^2},$$ we  have
\begin{align*}
\vert(Q_sh)_i\vert&\leq\left(\frac{n+s}{2}-1\right)\lambda_s^2\cdot\frac{(1+e^{2L})^2}{(n-1)e^{2L}}+\frac{(1+e^{2L})^2}{n(n-1)e^{2L}}s\left(\frac{n+s}{2}-1\right)\lambda_s^2 \\
&\leq\lambda_s^2\left(\frac{\frac{n+s}{2}-1}{n-1}e^{2L}+\frac{s\left(\frac{n+s}{2}-1\right)}{n(n-1)}e^{2L}\right)\tag{A.2.5}\label{A.2.5}
\end{align*}
Moreover, we have $\lambda_s=o_p(1)$, because it describes the consistency between MLE and the true parameter, which is gauranteed by regularity conditions in Mathematical Statistics, for exmaple, the Fisher information matrix is positive definite and derivatives can be passed under the integral sign. All these regularity can be satisfied because the logistic model is contained in Exponential Family and the Fisher information matrix equals to covariance matrix. So that by results in \ref{A.2.4} and \ref{A.2.5}, we have \[M_s^{-1}h=o_p(1)\tag{A.2.6}\label{A.2.6}\] as $n\rightarrow\infty,$ which finishes the claim.

Finally, combining \ref{A.2.1}, \ref{A.2.2}, \ref{A.2.3} and \ref{A.2.6}, we have
\begin{align*}
\Vert\hat \bsbeta_S-\bsbeta^*_S\Vert_2 & \leq s\left[O\left(\frac{e^{6L}}{(n-1)^2}\right)+\max_{1\leq i,j\leq s}\left\vert\frac{\delta_{ij}}{m_{ii}}-\frac{1}{m_{\cdot\cdot}}\right\vert\right]\cdot O_p\left(\max_{1\leq i\leq s}\sqrt{m_{ii}}\right) \\
& \leq s^{\frac{3}{2}}O_p(\sqrt n)\left[O\left(\frac{e^{6L}}{(n-1)^2}\right)+O\left(\frac{(1+e^{2L})^2}{ne^{2L}}\right)\right]\\
&\leq s^{\frac{3}{2}}O_p(\left(\frac{o((logs)^6)}{\sqrt n}\right),
\end{align*}under the assumption $L=o(loglogs).$ Then using triangle inequality\[\Vert\bsbeta^{m+1}-\bsbeta^*\Vert_2\leq\Vert\bsbeta^{m+1}-\hat \bsbeta\Vert_2+\Vert\hat \bsbeta-\bsbeta^*\Vert_2\] and results of \textbf{Theorem} \ref{Theorem 2.} finishes the proof.

\subsection{Proof of Theorem \ref{Theorem 3.3.}}\label{sec: GIC}

First, a generalized information criterion\citep{nishii1984asymptotic} is constructed as follows\[ \mathrm{GIC}^{*}_{a_n}(\mcalA):=\frac{2}{n(n-1)}\left(D\left(\hat \bsbeta^{ora}(\mcalA)\right)+a_n\vert\mcalA\vert\right), \]where $ D(\hat \bsbeta)=2(L_{sat}-L_n(\hat \bsbeta))$, with $L_{sat}$ the saturated model, $L_{n}(\bsbeta)=\sum_{i=1}^nd_i\beta_i-\sum_{1\leq i< j\leq n}log(1+e^{\beta_i+\beta_j})$ the positive log-likelihood function. Now we claim $a_n=O_p(logn)$ will identify the tuning parameter correctly.

By Proposition \ref{Proposition B.3.}, we have\[ \mathrm{GIC}^{*}_{a_n}(\mcalA)-\mathrm{GIC}^{*}_{a_n}(\mcalA^*)=\frac{4}{n(n-1)}I(\bsbeta^{KL}(\mcalA))+(\vert\mcalA\vert-\vert\mcalA^*\vert)\left(a_n\left(\frac{n(n-1)}{2}\right)^{-1} -O_p(R_n)\right), \]where $R_n=\frac{\sqrt{logn}}{n(n-1)}$. Then under the assumption in Theorem \ref{Theorem 3.3.}: $\delta_n K^{-1}R_n^{-1}\rightarrow\infty$ and $a_n=o(s^{-1}n(n-1)\delta_n)$ is chosen, for $n$ large enough, we have\[ \inf_{\mcalA\not\supset\mcalA^*,\vert\mcalA\vert\leq K}\mathrm{GIC}_{a_n}^{*}(\mcalA)-\mathrm{GIC}_{a_n}^{*}(\mcalA^*)\geq\delta_n/2 \tag{A.3.1}\label{A.3.1}, \]with probability tending to 1, where $\delta_n$ and $I(\bsbeta)$ is defined as \eqref{eq:minimum-kl-gap} and \eqref{eq:kl-divergence} correspondingly.

By Proposition \ref{Proposition B.4.}, we have uniformly for any model $\mcalA\not\supset\mcalA^*$ \[ \mathrm{GIC}_{a_n}^{*}(\mcalA)-\mathrm{GIC}^{*}_{a_n}(\mcalA^*)=\frac{2(\vert\mcalA\vert-\vert\mcalA^*\vert)}{n(n-1)}\left(a_n-O_p(\psi_n)\right)>\frac{a_n}{n(n-1)} ,\tag{A.3.2}\label{A.3.2} \] with asymptotic probability 1, if $a_n$ is chosen such that $a_n\psi_n^{-1}\rightarrow\infty$ with $\psi_n=logn$ defined as Proposition \ref{Proposition B.4.}.

Finally, under the choice of $a_n=logn\cdot log(logn)$, combining \ref{A.3.1} and \ref{A.3.2}, we have\[ P\left(\inf_{\mcalA\supsetneqq\mcalA^*} \mathrm{GIC}^{*}_{a_n}(\mcalA)-\mathrm{GIC}^{*}_{a_n}(\mcalA^*)\geq\frac{\delta_n}{2} \ \mathrm{and} \ \inf_{\mcalA\not\supset\mcalA^*} \mathrm{GIC}^{*}_{a_n}(\mcalA)-\mathrm{GIC}^{*}_{a_n}(\mcalA^*)>\frac{a_n}{n(n-1)}\right)\rightarrow 1 ,\] if $\delta_nK^{-1}R_n^{-1}\rightarrow\infty$ and $a_n$ satisfies $\frac{n(n-1)}{2}\delta_ns^{-1}a_n\rightarrow\infty$. This means that\[ P\left(\inf_{\mcalA\neq\mcalA^*}\mathrm{GIC}^{*}_{a_n}(\mcalA)>\mathrm{GIC}^{*}_{a_n}(\mcalA^*)\right)\rightarrow1, \]i.e., $a_n$ we chose can identify the correct tuning parameter accurately. Then omitting the coefficient of $\mathrm{GIC}^{*}_{a_n}$ and the saturated likelihood, we choose $\mathrm{GIC}^{*}=2\ell_n\left(\hat \bsbeta^{ora}(\mcalA)\right)+s\cdot log\left(\frac{n(n-1)}{2}\right)\cdot log(logn)$, which finishes the proof.

\subsection{Proof of Corollary \ref{Corollary 1.}}

This corollary is the direct result of Theorem \ref{Theorem 2.} and Theorem \ref{Theorem 3.3.}.

\section{Proofs of Lemmas}\label{App-B}

We give the proof of the property of surrogate function constructed above.

\subsection{Proof of Lemma \ref{Lemma 1.}}

\begin{lemma}\label{Lemma 5.}
A diagonal-dominant symmetric matrix whose diagonal entries are non-negative is positive semi-definite.
\end{lemma}
The lemma mentioned above can be easily found in some textbooks related to linear algebra. Now with this lemma at hand we can give the proof of Lemma \ref{Lemma 1.}.

By simple calculation \[ \frac{\partial^2\ell_n}{\partial\beta_i\partial\beta_j}=p_{ij}(1-p_{ij})\quad\forall i\neq j\qquad\frac{\partial^2\ell_n}{\partial\beta_i^2}=\sum_{j\neq i}p_{ij}(1-p_{ij}). \]

Let $T=diag(\boldsymbol{t}),\boldsymbol{t}=(t_1,\cdots,t_n)^T$, thus we need
\begin{itemize}
\item $\left\vert t_i-\sum_{j\neq i}p_{ij}(1-p_{ij})\right\vert\geq\sum_{j\neq i}\left\vert p_{ij}(1-p_{ij})\right\vert$,
\item $t_i\geq\sum_{j\neq i}p_{ij}(1-p_{ij})$.
\end{itemize}
Namely, $t_i\geq2\sum_{j\neq i}p_{ij}(1-p_{ij})$, observing that $2\sum_{j\neq i}p_{ij}(1-p_{ij})\leq\frac{n-1}{2}$, leading to $T=\frac{n-1}{2}I$ which is our surrogate Hesse matrix, i.e., $H\preceq\frac{n-1}{2}I$.

Now we give the proofs of the Lemmas in sections concerning \GICstar.

\subsection{Proof of Proposition B.1.}
\begin{proposition}\label{Proposition B.1.}
Under condition \ref{cond3.1}, \[ \sup_{\vert\mcalA\vert\leq K,\mcalA\subset[n]}\frac{2}{n(n-1)\vert\mcalA\vert}\left\vert L_n(\bsbeta^{KL}(\mcalA))-\bbE[L_n(\bsbeta^{KL}(\mcalA))]\right\vert=O_p\left(\frac{\sqrt{logn}}{n(n-1)}\right) .\]
\end{proposition}

We mainly use the Hoeffding inequality which can be find in \citep{buhlmann2011statistics} for the proof of Proposition B.1.

By definition, $L_n(\bsbeta^{KL}(\mcalA))-\bbE[L_n(\bsbeta^{KL}(\mcalA))]=\sum_{i=1}^n(d_i-\bbE[d_i])\beta_i^{KL}:=W^\top\bsbeta^{KL}$, where $W=(d_1-\bbE[d_1],\cdots,d_n-\bbE[d_n])^\top$, and shorthand $\bsbeta^*(\alpha)\equiv\bsbeta^{KL}$. By Hoeffding inequality, we have\[ P\left(\vert W^\top\bsbeta^{KL}\vert\geq t\right)\leq Cexp\left(\frac{-Ct^2}{\sum_{i=1}^n(\beta^{KL}_i)^2}\right) \]for some constant $C$. We set $t=\vert\alpha\vert\sqrt{\gamma_nlogn}$ with $\gamma_n$ some slowly diverging sequence and under the condition \ref{cond3.2}: $\Vert\bsbeta^{KL}(\mcalA)\Vert_2^2\leq C\vert\mcalA\vert$. Therefore, we have \[ \frac{1}{\vert\mcalA\vert}\left[L_n(\bsbeta^{KL}(\mcalA))-\bbE[L_n(\bsbeta^{KL}(\mcalA))]\right]=O_p\left(\sqrt{logn}\right), \] which finishes the proof.

\subsection{Proof of Proposition B.2.}
\begin{proposition}\label{Proposition B.2.}
Under condition \ref{cond3.1}, \[ \sup_{\vert\mcalA\vert\leq K,\mcalA\subset[n]}\frac{2}{n(n-1)\vert\mcalA\vert}\left\vert L_n(\hat\bsbeta^{ora}(\mcalA))-L_n(\bsbeta^{KL}(\mcalA))\right\vert=O_p\left(\frac{\sqrt{logn}}{n(n-1)}\right). \]
\end{proposition}

By definition, the following two points are satisfied:
\begin{enumerate}
\item For each given model $\mcalA$, we have \[L_n(\hat \bsbeta^{ora}(\mcalA))\geq L_n(\bsbeta^{KL}(\mcalA))\tag{B.3.1}\label{B.3.1}.\]
\item \[\bbE[L_n(\bsbeta^{KL}(\mcalA))-L_n(\hat \bsbeta^{ora}(\mcalA))]=I(\hat \bsbeta^{ora}(\mcalA))-I(\bsbeta^{KL}(\mcalA))\geq 0,\tag{B.3.2}\label{B.3.2}\] because $\bsbeta^{KL}(\mcalA)$ is the minimizer.
\end{enumerate}

Therefore, we have 
\begin{align*}
0&\overset{\ref{B.3.1}}{\leq} L_n(\hat \bsbeta^{ora}(\mcalA))-L_n(\bsbeta^{KL}(\mcalA)) \\
&=\left(L_n(\hat \bsbeta^{ora}(\mcalA))-\bbE[L_n(\hat \bsbeta^{ora}(\mcalA))]\right)+\bbE\left[L_n(\hat \bsbeta^{ora}(\mcalA))-L_n(\bsbeta^{KL}(\mcalA))\right] \\
&\quad+\left(\bbE[L_n(\bsbeta^{KL}(\mcalA))]-L_n(\bsbeta^{KL}(\alpha))\right) \\
&\overset{\ref{B.3.2}}{\leq}\left(L_n(\hat \bsbeta^{ora}(\mcalA))-\bbE[L_n(\hat \bsbeta^{ora}(\mcalA))]\right)-\left(L_n(\bsbeta^{KL}(\mcalA))-\bbE[L_n(\bsbeta^{KL}(\mcalA))]\right). 
\end{align*}
 And under the condition \ref{cond3.2} and Proposition \ref{Proposition B.1.}, we finish the proof.

\subsection{Proof of Proposition B.3.}
\begin{proposition}\label{Proposition B.3.}
Under conditions \ref{cond3.1}-\ref{cond3.2}, as $n\rightarrow\infty$, \[ \sup_{\vert\mcalA\vert\leq K,\mcalA\subset[n]}\frac{2}{n(n-1)\vert\mcalA\vert}\left\vert D(\hat \bsbeta^{ora}(\mcalA))-D(\hat \bsbeta^{ora}(\mcalA^*))-2I(\bsbeta^{KL}(\mcalA))\right\vert=O_p(R_n) ,\] with $R_n=\frac{\sqrt{logn}}{n(n-1)}$.
\end{proposition}
Proof:

By definition, $2I(\bsbeta^{KL}(\mcalA))=\bbE[D(\bsbeta^{KL}(\mcalA))-D(\bsbeta^*)]$. Thus, using the triangle inequality, we have
\begin{align*}
\vert D(\hat{\bsbeta}^{ora}&(\mcalA))-D(\hat \bsbeta^{ora}(\mcalA^*))-2I(\bsbeta^{KL}(\mcalA))\vert \\
&\leq\vert D(\hat\bsbeta^{ora}(\mcalA))-D(\bsbeta^{KL}(\mcalA))\vert+\vert D(\bsbeta^{KL}(\mcalA))-\bbE[D(\bsbeta^{KL}(\mcalA))]\vert \\
&+\vert  D(\hat\bsbeta^{ora}(\mcalA^*))-D(\bsbeta^{KL}(\mcalA^*))\vert+\vert D(\bsbeta^{KL}(\mcalA^*))-D(\bsbeta^*)\vert+\vert D(\bsbeta^*)-\bbE[D(\bsbeta^*)]\vert .
\end{align*}Noting that $\bsbeta^{KL}(\mcalA)=\bsbeta^*$ for all $\mcalA\supset\mcalA^*$, the last but one term is 0.

By Proposition \ref{Proposition B.1.}, we have 
\begin{gather*}
\frac{2}{n(n-1)\vert\mcalA\vert}\left\vert D(\bsbeta^{KL}(\mcalA))-\bbE[D(\bsbeta^{KL}(\mcalA))]\right\vert=O_p\left(\frac{\sqrt{logn}}{n(n-1)}\right);\\
\frac{2}{n(n-1)\vert\mcalA\vert}\left\vert D(\bsbeta^*)-\bbE[D(\bsbeta^*)]\right\vert=O_p\left(\frac{\sqrt{logn}}{n(n-1)}\right).
\end{gather*}

By Proposition \ref{Proposition B.2.}, we have 
\begin{gather*}
\frac{2}{n(n-1)\vert\mcalA\vert}\left\vert D(\hat\bsbeta^{ora}(\alpha))-D(\bsbeta^{KL}(\alpha))\right\vert=O_p\left(\frac{\sqrt{logn}}{n(n-1)}\right);\\
\frac{2}{n(n-1)\vert\mcalA\vert}\left\vert D(\hat\bsbeta^{ora}(\mcalA^*))-D(\bsbeta^{KL}(\mcalA^*))\right\vert=O_p\left(\frac{\sqrt{logn}}{n(n-1)}\right).
\end{gather*} Then we finish the proof.
\subsection{Proof of Proposition B.4.}
\begin{proposition}\label{Proposition B.4.}
Under conditions \ref{cond3.1}-\ref{cond3.2}, uniformly for all $\mcalA\supsetneqq\mcalA^*$ with $\vert\mcalA\vert\leq K$, as $n\rightarrow\infty$, \[ \frac{1}{\vert\mcalA\vert-\vert\mcalA^*\vert}\left(D(\hat \bsbeta^{ora}(\mcalA))-D(\hat \bsbeta^{ora}(\mcalA^*))\right)=O_p(\psi_n), \] with $\psi_n=logn$.
\end{proposition}

Proof:

Without loss of generality, we assume that $\mcalA^*=\{1,\cdots,s\},\mcalA=\{1,\cdots,t\}$ with $t>s$. By the definition of $\hat{\bsbeta}^{ora}(\mcalA)$ and $\bsbeta^{KL}(\mcalA)$, we have \[
d_i=\sum_{j\neq i}\frac{exp(\hat \beta_i^{ora}(\mcalA)+\hat \beta_j^{ora}(\mcalA))}{1+exp(\hat \beta_i^{ora}(\mcalA)+\hat \beta_j^{ora}(\mcalA))}\forall\ i=1,\cdots,t \ \mathrm{and} \ \hat \beta_j^{ora}(\mcalA)=\beta_0\quad\forall \ j=t+1,\cdots,n; \tag{B.4.(i)}\label{B.4.(i)} \] and \[
\sum_{j\neq i}\frac{exp(\beta^*_i+\beta^*_j)}{1+exp(\beta_i^*+\beta_j^*)}=\sum_{j\neq i}\frac{exp(\beta_i^{KL}(\mcalA)+\beta_j^{KL}(\mcalA))}{1+exp(\beta_i^{KL}(\mcalA)+\beta_j^{KL}(\mcalA))}\forall \ i=1,\cdots,t \ \mathrm{and} \ \beta_j^{KL}(\mcalA)=\beta_0\quad\forall \ j=t+1,\cdots,n\tag{B.4.(ii)}\label{B.4.(ii)}
.\] and the second equation is the result of the derivative of the KL-divergence \eqref{eq:kl-divergence}.

Moreover, we have $\beta^*_i=\beta_0\quad\forall i=s+1,\cdots,n$ by definition. We always refer to $\beta_0$ as an unknown but a constant parameter.

By directly calculate, we find $D(\hat \bsbeta^{ora}(\mcalA))-D(\hat \bsbeta^{ora}(\mcalA^*))=2\left(L_n(\hat \bsbeta^{ora}(\mcalA^*)-L_n(\hat \bsbeta^{ora}(\mcalA))\right)$. Still by using $\bsbeta^{KL}(\mcalA)=\bsbeta^{KL}(\mcalA^*)=\bsbeta^*$ for $\mcalA^*\subsetneqq\mcalA$, we have \[ L_n(\hat \bsbeta^{ora}(\mcalA^*))-L_n(\hat \bsbeta^{ora}(\mcalA))=\left(L_n(\hat \bsbeta^{ora}(\mcalA^*))-L_n(\bsbeta^{KL}(\mcalA^*))\right)-\left(L_n(\hat \bsbeta^{ora}(\mcalA))-L_n(\bsbeta^{KL}(\mcalA))\right). \tag{B.4.1}\label{B.4.1}\]

From \ref{B.4.1} above, we only need to make an estimation for $L_n(\hat \bsbeta^{ora}(\mcalA))-L_n(\bsbeta^{KL}(\mcalA))$ for $\mcalA\supset\mcalA^*$.

By Taylor expansion, we have  
\[ L_n(\hat \bsbeta^{ora}(\mcalA))-L_n(\bsbeta^{KL}(\mcalA))=(G^{KL})^\top(\hat \bsbeta^{ora}(\mcalA)-\bsbeta^{KL}(\mcalA))+\frac{1}{2}(\hat{\bsbeta}^{ora}(\mcalA)-\bsbeta^{KL}(\mcalA))^\top\tilde H(\hat \bsbeta^{ora}(\mcalA)-\bsbeta^{KL}(\mcalA)), \]where $G^{KL}$ is evaluated at $\bsbeta^{KL}(\mcalA)$ and $\tilde H$ is evaluated at $\tilde{\bsbeta}$ lying in the line segment of $\hat \bsbeta^{ora}(\mcalA)$ and $\bsbeta^{KL}(\mcalA)$.

 Using the conclusion of Lemma \ref{Lemma 1.} about the partial order of the Hesse matrix, we have\[ \left\vert L_n(\hat \bsbeta^{ora}(\mcalA))-L_n(\bsbeta^{KL}(\mcalA))\right\vert\leq\Vert G^{KL}\Vert_2\Vert\hat{\bsbeta}^{ora}(\mcalA)-\bsbeta^{KL}(\mcalA)\Vert_2+\frac{n-1}{4}\Vert\hat \bsbeta^{ora}(\mcalA)-\bsbeta^{KL}(\mcalA)\Vert^2_2. \tag{B.4.2}\label{B.4.2}\]
 
Using \ref{B.4.(i)} and \ref{B.4.(ii)}, when $i=1,\cdots, t$, i.e., $i\in\mcalA$ , we have 
\begin{align*}
G_i^{KL}&=d_i-\sum_{j\neq i}\frac{exp(\beta_i^{KL}(\mcalA)+\beta_j^{KL}(\mcalA))}{1+exp(\beta_i^{KL}(\mcalA)+\beta_j^{KL}(\mcalA))} \\
&=\sum_{j\neq i}\left[\frac{exp(\hat \beta_i^{ora}(\mcalA)+\hat \beta_j^{ora}(\mcalA))}{1+exp(\hat \beta_i^{ora}(\mcalA)+\hat \beta_j^{ora}(\mcalA))}-\frac{exp(\beta_i^{KL}(\mcalA)+\beta_j^{KL}(\mcalA))}{1+exp(\beta_i^{KL}(\mcalA)+\beta_j^{KL}(\mcalA))}\right] \\
&=\sum_{j\in\mcalA,j\neq i}\left[\frac{exp(\hat \beta_i^{ora}(\mcalA)+\hat \beta_j^{ora}(\mcalA))}{1+exp(\hat \beta_i^{ora}(\mcalA)+\hat \beta_j^{ora}(\mcalA))}-\frac{exp(\beta_i^{KL}(\mcalA)+\beta_j^{KL}(\mcalA))}{1+exp(\beta_i^{KL}(\mcalA)+\beta_j^{KL}(\mcalA))}\right] \\
&+\sum_{j\in[n]\backslash\mcalA}\left[\frac{exp(\hat \beta_i^{ora}(\mcalA)+\beta_0)}{1+exp(\hat \beta_i^{ora}(\mcalA)+\beta_0)}-\frac{exp(\beta_i^{KL}(\mcalA)+\beta_0)}{1+exp(\beta_i^{KL}(\mcalA)+\beta_0)}\right] \\
&=C_1\sum_{j\in\mcalA,j\neq i}\left[\hat \beta_i^{ora}(\mcalA)+\hat\beta_j^{ora}(\mcalA)-\beta_i^{KL}(\mcalA)-\beta_j^{KL}(\mcalA)\right]+C_2\sum_{j\in[n]\backslash\mcalA}\left[\hat \beta_i^{ora}(\mcalA)-\beta_i^{KL}(\mcalA)\right] \\
&=C_1\left[(t-1)(\hat \beta_i^{ora}(\mcalA)-\beta_i^{KL}(\mcalA))+\sum_{j\in\mcalA,j\neq i}(\hat \beta_j^{ora}(\mcalA)-\beta_j^{KL}(\mcalA))\right]+C_2(n-t)(\hat \beta_i^{ora}(\mcalA)-\beta_i^{KL}(\mcalA)),
\end{align*}where the technique of constants $C_1,C_2$ appears again since the proof of Theorem \ref{Theorem 2.}. Therefore, we have 
\begin{align*}
G_i^{KL}(\hat \beta_i^{ora}(\mcalA)-\beta_i^{KL}(\mcalA))&=C_1(t-1)(\hat \beta_i^{ora}(\mcalA)-\beta_i^{KL}(\mcalA))^2+C_2(n-t)(\hat \beta_i^{ora}(\mcalA)-\beta_i^{KL}(\mcalA))^2 \\
&+C_1(\hat \beta_i^{ora}(\mcalA)-\beta_i^{KL}(\mcalA))\sum_{j\in\mcalA,j\neq i}(\hat \beta_j^{ora}(\mcalA)-\beta_j^{KL}(\mcalA)) \\
&=\left[C_1(t-1)+C_2(n-t)\right](\hat \beta_i^{ora}(\mcalA)-\beta_i^{KL}(\mcalA))^2 \\
&+C_1(\hat \beta_i^{ora}(\mcalA)-\beta_i^{KL}(\mcalA))\sum_{j\in\mcalA,j\neq i}(\hat \beta_j^{ora}(\mcalA)-\beta_j^{KL}(\mcalA)).
\end{align*}
Thus, 
\begin{align*}
(G^{KL})^\top&(\hat \bsbeta^{ora}(\mcalA)-\bsbeta^{KL}(\mcalA)) \tag{B.4.3}\label{B.4.3}\\
&\leq\sum_{i=1}^t\bigg\{\left[C_1(t-1)+C_2(n-t)\right](\hat{\beta}_i^{ora}(\mcalA)-\beta_i^{KL}(\mcalA))^2 \\
&+C_1(\hat \beta_i^{ora}(\mcalA)-\beta_i^{KL}(\mcalA))\sum_{j\in\mcalA,j\neq i}(\hat \beta_j^{ora}(\mcalA)-\beta_j^{KL}(\mcalA))\bigg\}.
\end{align*}

Observing that
\begin{align*}
\sum_{i=1}^t(\hat \beta_i^{ora}(\mcalA)-\beta_i^{KL}(\mcalA))&\sum_{j\in\mcalA,j\neq i}(\hat \beta_j^{ora}(\mcalA)-\beta_j^{KL}(\mcalA)) \\
&\leq\frac{1}{2}\sum_{i=1}^t\sum_{j\in\mcalA,j\neq i}\left[(\hat \beta_i^{ora}(\mcalA)-\beta_i^{KL}(\mcalA))^2+(\hat \beta_j^{ora}(\mcalA)-\beta_j^{KL}(\mcalA))^2\right] \\
&=\frac{1}{2}\left\{\sum_{i=1}^t(t-1)(\hat \beta_i^{ora}(\mcalA)-\beta_i^{KL}(\mcalA))^2+\sum_{i=1}^t\sum_{j\in\mcalA,j\neq i}(\hat \beta_j^{ora}(\mcalA)-\beta_j^{KL}(\mcalA))^2\right\} \\
&=\sum_{i=1}^t(t-1)(\hat \beta_i^{ora}(\mcalA)-\beta_i^{KL}(\mcalA))^2,
\end{align*}
where the inequality is the result of $ab\leq\frac{a^2+b^2}{2}$.
Plugging the term above into \ref{B.4.3}, we have 
\begin{align*}
\vert (G^{KL})^\top(\hat \bsbeta^{ora}(\mcalA)-\bsbeta^{KL}(\mcalA))\vert&\leq\left(C_1(t-1)+C_2(n-t)+C_1(t-1)\right)\sum_{i=1}^{t}\left(\hat \beta_i^{ora}(\mcalA)-\beta_i^{KL}(\mcalA)\right)^2 \\
&\leq\frac{1}{4}(n+\vert\mcalA\vert-2)\Vert\hat \bsbeta^{ora}(\mcalA)-\bsbeta^{KL}(\mcalA)\Vert^2_2,
\end{align*} where $C_1,C_2\leq\frac{1}{4}$ by Lagrange mean value Theorem and surely 
\[
\Vert\hat \bsbeta^{ora}(\mcalA)-\bsbeta^{KL}(\mcalA)\Vert^2_2=\sum_{i=1}^{t}\left(\hat \beta_i^{ora}(\mcalA)-\beta_i^{KL}(\mcalA)\right)^2
.
\]

Combining \ref{B.4.2}, we obtain\[ \left\vert L_n(\hat \bsbeta^{ora}(\mcalA))-L_n(\bsbeta^{KL}(\mcalA))\right\vert\leq\frac{1}{4}\left(2n-2+\vert\mcalA\vert\right)\left\Vert\hat \bsbeta^{ora}(\mcalA)-\bsbeta^{KL}(\mcalA)\right\Vert^2_2. \tag{B.4.4}\label{B.4.4}\]

Now, we come to estimate $\left\Vert\hat \bsbeta^{ora}(\mcalA)-\bsbeta^{KL}(\mcalA)\right\Vert_2$. We first have the following relation
\begin{align*}
\vert\mcalA\vert& O_p\left(\sqrt{logn}\right)\overset{(i)}{=}\left(L_n(\hat \bsbeta^{ora}(\mcalA))-\bbE[L_n(\hat \bsbeta^{ora}(\mcalA))]\right)-\left(L_n(\bsbeta^{KL}(\mcalA))-\bbE[L_n(\bsbeta^{KL}(\mcalA))]\right) \\
&\overset{(ii)}{\geq}\bbE\left[L_n(\bsbeta^{KL}(\mcalA))-L_n(\hat \bsbeta^{ora}(\mcalA))\right]\overset{(iii)}{=}I(\hat \bsbeta^{ora}(\mcalA))-I(\bsbeta^{KL}(\mcalA))\overset{(iv)}{\geq}0. \tag{B.4.5}\label{B.4.5}
\end{align*}
\begin{enumerate}
\item (i) is the direct result from \ref{Proposition B.1.}.
\item (ii) uses $L_n(\hat \bsbeta^{ora}(\mcalA))-L_n(\bsbeta^{KL}(\mcalA))\geq0$ by definition.
\item (iii) is the result of direct calculation.
\item (iv) is the definition of $\bsbeta^{KL}(\mcalA)$.
\end{enumerate}

We solve the problem by evaluating\[ \bbE\left[L_n(\bsbeta^{KL}(\mcalA))-L_n(\hat \bsbeta^{ora}(\mcalA))\right]=\sum_{i=1}^t\bbE[d_i](\beta_i^{KL}(\mcalA)-\hat \beta_i^{ora}(\mcalA))+\sum_{1\leq i<j\leq n}log\left(1+exp(\hat \beta_i^{ora}(\mcalA)+\hat \beta_j^{ora}(\mcalA))\right)\] \[-log\left(1+exp(\beta_i^{KL}(\mcalA)+\beta_j^{KL}(\mcalA))\right). \tag{B.4.6}\label{B.4.6}\]

After then, we have the attention\[ \bbE[d_i]=\sum_{j\neq i}\frac{exp(\beta^*_i+\beta^*_j)}{1+exp(\beta^*_i+\beta^*_j)}\overset{\ref{B.4.(ii)}}{=}\sum_{j\neq i}\frac{exp(\beta_i^{KL}(\mcalA)+\beta_j^{KL}(\mcalA))}{1+exp(\beta_j^{KL}(\mcalA)+\beta_j^{KL}(\mcalA))}. \] Then by plugging into \ref{B.4.6} we have
\begin{align*}
\bbE\left[L_n(\bsbeta^{KL}(\mcalA))-L_n(\hat \bsbeta^{ora}(\mcalA))\right]&=\sum_{i=1}^t\sum_{j\neq i}\frac{exp(\beta_i^{KL}(\mcalA)+\beta_j^{KL}(\mcalA))}{1+exp(\beta_i^{KL}(\mcalA)+\beta_j^{KL}(\mcalA))}(\beta_i^{KL}(\mcalA)-\hat \beta_i^{KL}(\mcalA)) \\
+\sum_{1\leq i<j\leq n} & log\left(1+exp(\hat \beta_i^{ora}(\mcalA)+\hat \beta_j^{ora}(\mcalA))\right)-log\left(1+exp(\beta_i^{KL}(\mcalA)+\beta_j^{KL}(\mcalA))\right) \\
&:=\Gamma+\Theta, 
\end{align*}
where\[ \Gamma=\sum_{i=1}^t\sum_{j\neq i}\frac{exp(\beta_i^{KL}(\mcalA)+\beta_j^{KL}(\mcalA))}{1+exp(\beta_i^{KL}(\mcalA)+\beta_j^{KL}(\mcalA))}(\beta_i^{KL}(\mcalA)-\hat \beta_i^{KL}(\mcalA)), \] and\[ \Theta=\sum_{1\leq i<j\leq n}log\left(1+exp(\hat \beta_i^{ora}(\mcalA)+\hat \beta_j^{ora}(\mcalA))\right)-log\left(1+exp(\beta_i^{KL}(\mcalA)+\beta_j^{KL}(\mcalA))\right). \]
Using Lagrange mean value Theorem again, we make the following estimations:
\begin{itemize}
\item $\vert \Gamma\vert\leq\sum_{i=1}^t\sum_{j\neq i}\left\vert\beta_i^{KL}(\mcalA)-\hat \beta_i^{ora}(\mcalA)\right\vert=(n-1)\sum_{i=1}^t\left\vert\beta_i^{KL}(\mcalA)-\hat \beta_i^{ora}(\mcalA)\right\vert$ because of the boundedness of $\frac{e^x}{1+e^x}$.
\item $\vert \Theta\vert\leq C\sum_{1\leq i<j\leq n}\left\vert\hat \beta_i^{ora}(\mcalA)+\hat \beta_j^{ora}(\mcalA)-\beta_i^{KL}(\mcalA)-\beta_j^{KL}(\mcalA)\right\vert\leq C(n-1)\sum_{i=1}^t\left\vert\hat \beta_i^{ora}(\mcalA)-\beta_i^{KL}(\mcalA)\right\vert$ by using Lagrange mean value Theorem on the function $log(1+x)$ for some constant $C$.
\end{itemize}
Therefore, we know that 
\begin{align*}
&\bbE\left[L_n(\bsbeta^{KL}(\mcalA))-L_n(\hat \bsbeta^{ora}(\mcalA))\right]\\*
&=O_p(n)\cdot\left\Vert\hat \bsbeta^{ora}(\mcalA)-\bsbeta^{KL}(\mcalA)\right\Vert_1\\*
&\geq O_p(n)\left\Vert\hat \bsbeta^{ora}(\mcalA)-\bsbeta^{KL}(\mcalA)\right\Vert_2
.
\end{align*} Then we plug into \ref{B.4.5}, we have\[ \Vert\hat\bsbeta^{ora}(\mcalA)-\bsbeta^{KL}(\mcalA)\Vert_2\leq\vert\mcalA\vert O_p\left(\frac{\sqrt{logn}}{n}\right). \tag{B.4.7}\label{B.4.7}\]

Plugging \ref{B.4.7} into \ref{B.4.4}, we get\[ \left\vert L_n(\hat \bsbeta^{ora}(\mcalA))-L_n(\bsbeta^{KL}(\mcalA))\right\vert\leq\left(O_p(n)+\vert\mcalA\vert\right)\vert\mcalA\vert^2O_p\left(\frac{logn}{n^2}\right). \tag{B.4.8}\label{B.4.8}\]

Finally, plugging \ref{B.4.8} into \ref{B.4.1}, we obtain under the Condition \ref{cond3.1}( $\vert\mcalA\vert\leq K=O(\sqrt n)$): \[ \left\vert D(\hat \bsbeta^{ora}(\mcalA))-D(\bsbeta^{KL}(\mcalA))\right\vert=O_p(logn), \] because $\vert\mcalA\vert-\vert\mcalA^*\vert$ can be viewed as constant in practical situations.

\end{document}